\documentclass[onecolumn,aps,prd,preprintnumbers,superscriptaddress,nofootinbib,amsmath,amssymb,floats,floatfix,showkeys,notitlepage,showpacs]{revtex4-1}

\usepackage{orcidlink}
\usepackage{comment}
\usepackage{lipsum}
\usepackage{graphicx}
\usepackage{subfigure}
\usepackage{palatino}
\usepackage{sans}
\usepackage{hyperref}
\hypersetup{colorlinks=true,linkcolor=blue,urlcolor=blue,citecolor=blue}
\usepackage[toc,page]{appendix}
\usepackage[normalem]{ulem}
\usepackage{adjustbox}
\usepackage{latexsym}
\usepackage{amsmath}
\usepackage{amssymb}
\usepackage{amsfonts}
\usepackage{dcolumn}
\usepackage{bm}
\usepackage{tikz}
\usepackage{bigints}
\usepackage{array,tabularx,multirow,booktabs}
\usepackage[tracking=true]{microtype}
\SetTracking{}{500}
\SetTracking{encoding={*}, shape=sc}{40}
\UseRawInputEncoding 
\allowdisplaybreaks

\begin{document}
	\newcommand \nn{\nonumber}
	\newcommand \fc{\frac}
	\newcommand \lt{\left}
	\newcommand \rt{\right}
	\newcommand \pd{\partial}
	\newcommand \e{\text{e}}
	\newcommand \hmn{h_{\mu\nu}}
	\newcommand{\PR}[1]{\ensuremath{\left[#1\right]}} 
	\newcommand{\PC}[1]{\ensuremath{\left(#1\right)}} 
	\newcommand{\PX}[1]{\ensuremath{\left\lbrace#1\right\rbrace}} 
	\newcommand{\BR}[1]{\ensuremath{\left\langle#1\right\vert}} 
	\newcommand{\KT}[1]{\ensuremath{\left\vert#1\right\rangle}} 
	\newcommand{\MD}[1]{\ensuremath{\left\vert#1\right\vert}} 

	
\title{Testing black holes with cosmological constant in Einstein-bumblebee gravity through the black hole shadow using EHT data and deflection angle}

\author{Reggie C. Pantig \orcidlink{0000-0002-3101-8591}}
\email{rcpantig@mapua.edu.ph}
\affiliation{Physics Department, School of Foundational Studies and Education, Map\'ua University, 658 Muralla St., Intramuros, Manila 1002, Philippines.}

\author{
Shubham Kala
 \orcidlink{0000-0003-2379-0204}}
\email{shubhamkala871@gmail.com}
\affiliation{The Institute of Mathematical Sciences, C.I.T. Campus, Taramani, Chennai 600113, India.}

\author{Ali \"Ovg\"un \orcidlink{0000-0002-9889-342X}}
\email{ali.ovgun@emu.edu.tr}
\affiliation{Physics Department, Eastern Mediterranean University, Famagusta, 99628 North
Cyprus via Mersin 10, Turkiye.}

\author{Nikko John Leo S. Lobos \orcidlink{0000-0001-6976-8462}}
\email{nslobos@ust.edu.ph}
\affiliation{Electronics Engineering Department, University of Santo Tomas, Espa\~na Blvd, Sampaloc, Manila, 1008 Metro Manila, Philippines.}

\begin{abstract}
This study probes spacetime solutions within Einstein-Bumblebee gravity, a modified gravitational framework incorporating spontaneous Lorentz symmetry violation through a vector field mechanism. By introducing a cosmological constant into this model, the research scrutinizes thermodynamic properties of black holes in both anti-de Sitter (AdS) and de Sitter (dS) geometries. The investigation demonstrates how Lorentz-violating parameters alter foundational thermodynamic principles, including revisions to the first law of black hole mechanics and shifts in critical phenomena during phase transitions. Notably, the bumblebee coupling parameter emerges as a critical factor governing horizon structure and thermal emission characteristics, with pronounced deviations from general relativity (GR) predictions observed as this parameter increases. The analysis extends to observational signatures by calculating shadow profiles of these modified black holes. Shadow morphology exhibits dual dependence on the cosmological constant and the bumblebee parameter, presenting measurable discrepancies from classical relativity that could be constrained through Event Horizon Telescope (EHT) observational data. Furthermore, using geometric formalisms, the study quantifies light deflection phenomena in weak and strong gravitational regimes. Results reveal that both the cosmological constant and Lorentz-violating parameter induce detectable modifications to lensing angles compared to Schwarzschild or Kerr benchmarks. These deviations, while subtle, underscore the necessity for next-generation astronomical instruments capable of resolving fine-scale spacetime curvature effects.
\end{abstract}	



\maketitle


\section{Introduction}\label{intro}  
Recent experimental confirmations of black hole existence have corroborated numerous theoretical models developed in recent decades \cite{Einstein:1915ca,Wald:1984rg,chandrasekhar1998mathematical,LIGOScientific:2016aoc,LIGOScientific:2021qlt,EventHorizonTelescope:2022xqj}. Parallel advancements in understanding Lorentz symmetry violations in low-energy contexts, influenced by quantum gravitational phenomena at Planck-scale energies, have increasingly attracted scientific interest \cite{Liberati:2013xla}. Within this landscape, the Bumblebee model has emerged as a gravitational modification to explore potential symmetry-breaking effects on cosmological dynamics and gravitational interactions. This framework introduces spontaneous Lorentz symmetry breaking (LSB) via a vector field acquiring a non-zero vacuum expectation value aligned with a preferred spatial direction \cite{PhysRevD.63.065008,PhysRevD.39.683}. In Einstein-Bumblebee gravity, the synergy between the cosmological constant and Lorentz-violating parameters creates a richer theoretical environment \cite{Maluf:2020kgf} for probing black hole optical characteristics and thermal behavior. Analyzing these phenomena enables researchers to refine fundamental physics principles, test GR boundaries, and explore quantum gravity implications. The integration of the cosmological constant into Einstein-Bumblebee gravity aims to explore how GR modifications through dynamical vector fields could affect cosmic expansion dynamics. This unified approach facilitates novel interpretations of EHT observations by linking spacetime optical properties to cosmological parameters \cite{Islam:2024sph,Afrin:2024khy}.  

Black hole thermodynamics remains a captivating area of research in contemporary cosmology and astrophysics. Seminal contributions by Bekenstein and Hawking established crucial connections between black hole mechanics and thermodynamic principles \cite{bekenstein2020black,hawking1975particle}, laying groundwork for semi-classical thermodynamic analyses of GR-derived and modified-gravity black hole solutions. Investigations into AdS black hole thermodynamics gained prominence following Hawking's identification of phase transitions in Schwarzschild-AdS spacetimes \cite{Hawking:1982dh}. Subsequent breakthroughs by Chamblin revealed charge-dependent phase transitions in AdS black holes \cite{Chamblin:1999tk}, while Kastor et al. redefined AdS black hole mass as enthalpy within extended phase space thermodynamics, associating cosmological constants with thermodynamic pressure \cite{Kastor:2009wy}. These developments spurred extensive studies on equations of state for rotating and higher-dimensional black holes \cite{Banerjee:2011au,Gunasekaran:2012dq,Zou:2013owa,Dolan:2014lea,Promsiri:2020jga,Walia:2021emv,Ali:2023ppg,Cano:2024tcr}, with Kubizňák et al. providing comprehensive analyses of charged AdS black hole criticality \cite{kubizvnak2012p}. Recent explorations span diverse spacetime geometries \cite{Haditale:2023adr}, massive gravity frameworks \cite{jafarzade2024thermodynamics}, and loop quantum gravity corrections \cite{wang2024thermodynamics}, underscoring the importance of thermodynamic investigations in Einstein-Bumblebee modified black holes.  

A black hole shadow manifests as a dark profile observed against luminous surrounding matter, produced through extreme light bending near the event horizon. Synge first quantified photon trajectories near Schwarzschild black holes \cite{Synge:1966okc}, with Bardeen later formalizing shadow characteristics for distant observers \cite{Zakharov:2023yjl}. Subsequent decades refined theoretical foundations \cite{Luminet:1979nyg, Falcke:1999pj, Claudel:2000yi,Cunningham,1974IAUS...64..132B}, while modern advances in numerical relativity and interferometric techniques enabled the groundbreaking EHT imaging of M87* and Sgr A* \cite{EventHorizonTelescope:2019dse, EventHorizonTelescope:2019ths, EventHorizonTelescope:2022xqj}. Shadow morphology encodes critical spacetime geometry information, permitting parameter estimation and GR validity tests \cite{Zakharov:2011zz}. Recent studies extensively analyze shadow dependencies on black hole charge \cite{Zakharov:2005ek}, spin \cite{EventHorizonTelescope:2021dqv}, and alternative gravity parameters \cite{Gao:2024ejs,Lambiase:2024uzy, Li:2024owp, Atamurotov:2024nre,Chen:2024mlr,Papnoi:2024wzs,Yildiz:2024dkt,Zare:2024dtf}, establishing shadows as key observational probes for modified gravity theories.  

Gravitational lensing, a cornerstone prediction of GR, describes light deflection near massive celestial bodies. Early theoretical frameworks by Refsdal \cite{Refsdal:1964yk} and Virbhadra \cite{PhysRevD.62.084003} established relativistic image formation principles, while Gibbons-Werner methods utilizing Gauss-Bonnet topology \cite{Gibbons_2008} revolutionized deflection angle calculations in curved spacetimes. Subsequent extensions to stationary geometries \cite{Werner:2012rc} and finite-distance scenarios \cite{PhysRevD.94.084015,PhysRevD.95.044017} enhanced lensing analysis precision. Modern applications employ these geometric techniques across diverse spacetime configurations \cite{PhysRevD.95.104012,PhysRevD.97.024042,PhysRevD.98.044033,PhysRevD.101.024040}, providing insights into compact object properties and gravitational theory validation.  

Strong-field gravitational lensing near photon spheres exhibits diverging deflection angles, as demonstrated by Bozza's systematic approach \cite{bozza2002gravitational}. Subsequent refinements by Tsukamoto \cite{tsukamoto2017deflection} extended methodology applicability while preserving core predictions. Mushtaq et al.~\cite{Mushtaq:2024tmp} investigate the deflection angle of light and the characteristics of quasi-periodic oscillations in an extended gravitationally decoupled black hole spacetime, revealing novel signatures that could distinguish such solutions observationally. Mustafa et al.~\cite{Mustafa:2024mvx} analyze epicyclic oscillation frequencies around slowly rotating charged black holes in the context of Bumblebee gravity, demonstrating modifications to the radial and vertical modes induced by LSB. Wang and Battista~\cite{Wang:2025fmz} study dynamical trajectories and shadow features of a quantum-corrected Schwarzschild black hole within effective field theories of gravity, showing that quantum corrections can lead to observable deviations in the shadow radius. Capozziello et al.~\cite{Capozziello:2024ucm} propose a mechanism based on atemporality to avoid curvature singularities in Lorentzian–Euclidean black hole models, offering a novel pathway to nonsingular gravitational solutions. Sekhmani et al.~\cite{Sekhmani:2025kav} explore thermodynamic behavior and phase transitions of AdS black holes coupled to ModMax nonlinear electrodynamics and perfect fluid dark matter, uncovering rich critical phenomena and modified equation of state parameters. Javed et al.~\cite{Javed:2024nnt} examine the Joule-Thomson expansion of charged AdS black holes with nonlinear electrodynamics and thermal fluctuations via Barrow entropy, identifying conditions under which cooling-heating inversion points occur. Ashraf et al.~\cite{Ashraf:2025edl} test rotating Einstein-Yang-Mills-Higgs black hole models through their predicted quasi-periodic oscillations, finding distinctive frequency spectra that could serve as observational probes of the underlying gauge fields. Recent applications to Einstein-Bumblebee gravity models \cite{poulis2022exact} reveal measurable deviations from GR predictions, highlighting lensing's potential for probing LSB effects \cite{Kostelecky:2020hbb,Bertolami:2005bh,Casana:2017jkc}. These investigations complement thermodynamic and shadow analyses, collectively constraining modified gravity parameters through multi-messenger astrophysical observations.  

This work employs geometrized units ($G=c=1$) throughout. Section \ref{sec2} derives cosmological constant-embedded black hole solutions in Einstein-Bumblebee gravity, analyzing horizon structures. Section \ref{sec3} investigates thermodynamic properties, including extended phase space formalism and critical phenomena. Section \ref{sec4} computes shadow characteristics constrained by EHT observations. Sections \ref{sec5} and \ref{sec6} analyze weak and strong deflection angle (SDA) via geometric methods. Section \ref{conc} synthesizes key findings and implications for modified gravity research.  

\section{Black hole with cosmological constant spacetime in Einstein-bumblebee gravity} \label{sec2}
The metric for an asymptotically flat black hole solution in Einstein-Bumblebee gravity is expressed as follows \cite{Maluf:2020kgf}, 
\begin{equation}
    ds^{2} = A(\rho) dt^{2} - \frac{1+\ell}{A(\rho)} d \rho^{2} - \rho^{2} d\Omega^{2}.
\end{equation}
Here,
\begin{equation} \label{coeff}
    A(\rho) = 1-\frac{2M}{\rho}-(1+\ell)\frac{\Lambda_{e}}{3}\rho^{2},
    \end{equation}
with $\Lambda_{e}=k\lambda/Q$ denoting the effective cosmological constant. $M$ represents the mass of black hole and $l$ is the bumblebee parameter.
The horizon of a black hole can be obtain by solving, $ A(\rho) = 1-2M/\rho-(1+\ell)\frac{\Lambda_{e}}{3}\rho^{2}=0$, or equivalently the cubic equation,
\begin{equation}
    \rho^{3} - \frac{3\rho}{\Lambda_{e}(1+l)} + \frac{6M}{\Lambda_{e}(1+\ell)} =0.
\end{equation}
The solution of this cubic equation can be obtained by the usual Cardan-Tartaglia method. The solution can be classified based on the value of the effective cosmological constant and the range of black hole parameters.
\begin{figure}[h]
	\begin{center}
        {\includegraphics[width=0.48\textwidth]{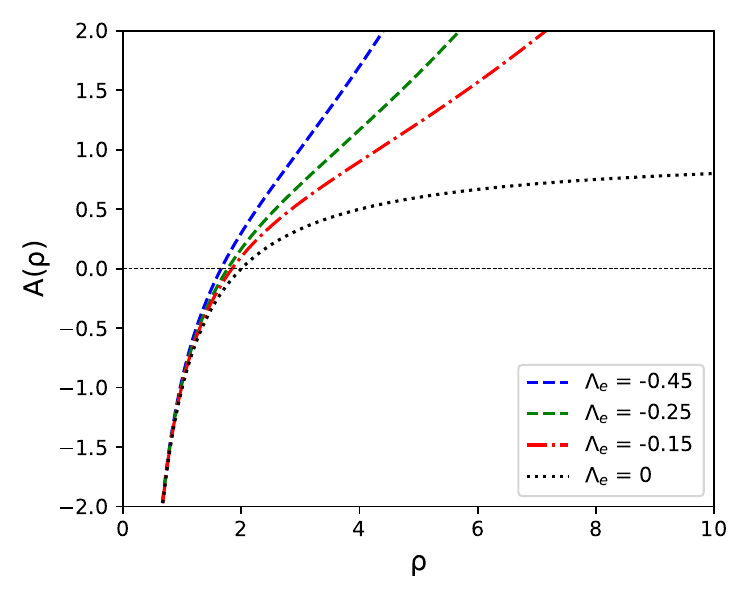}} 
        {\includegraphics[width=0.48\textwidth]{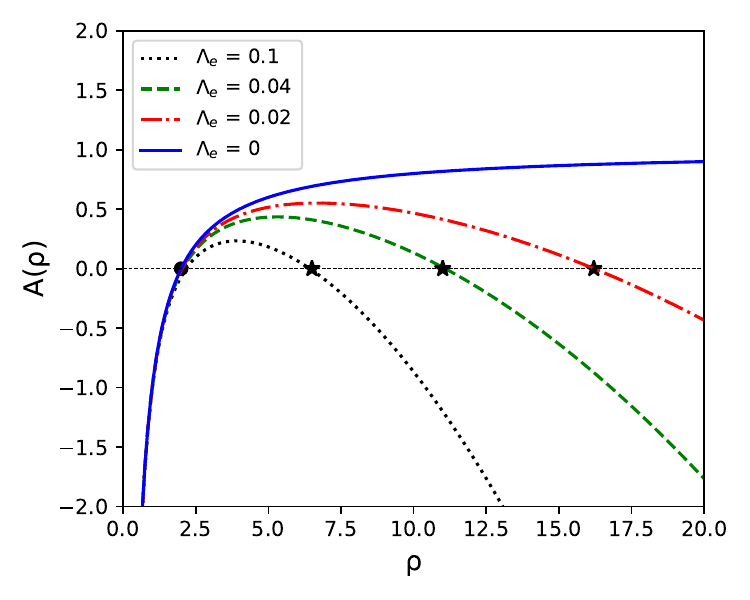}}
	\end{center}
    \caption{Plot of $ A(\rho) $ versus radial distance $\rho$ for AdS (left panel, $\Lambda_e = -0.1$) and dS (right panel, $\Lambda_e = 0.1$), with $ M = 1 $ and $\ell = 0.5 $.}
    \label{structure Horizon}
\end{figure}

The graphical representation of $A(\rho)=0$ is illustrated in \figurename{\ref{structure Horizon}} to understand the nature of the black hole horizon. It is observed that in the case of AdS symmetry, there exists a single horizon. However, for the dS case, two horizons exist. The position of the black horizon, represented by a small black circle, and the cosmological horizons are depicted by a black star, respectively. In particular, it has been observed that the radius of the cosmological horizon increased with a decrease in $\Lambda_{e}$. \textcolor{black}{The selection of specific $\Lambda$ values is guided by the underlying metric parameters, ensuring a consistent representation of AdS ($\Lambda < 0$) and dS ($\Lambda > 0$) spacetime geometries. Since we are using geometrized units, the cosmological constant has units of inverse length squared.}

\begin{figure}[h]
	\begin{center}
        {\includegraphics[width=0.48\textwidth]{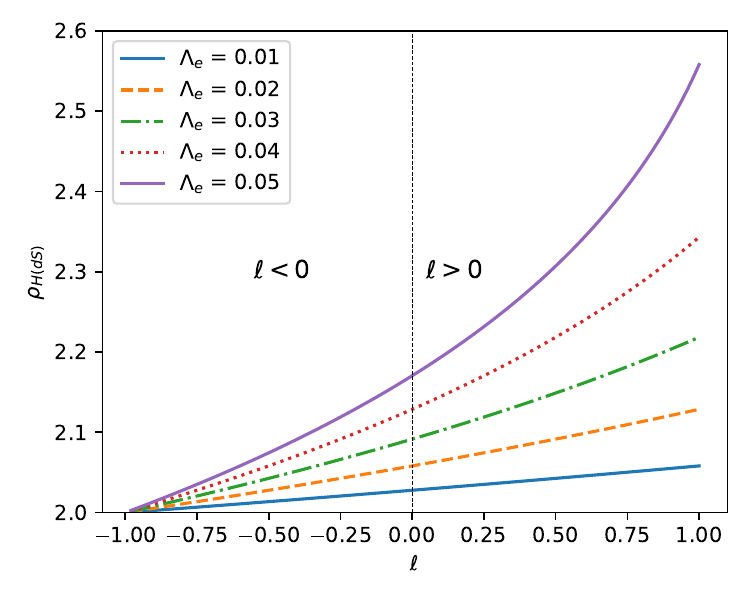}} 
        {\includegraphics[width=0.48\textwidth]{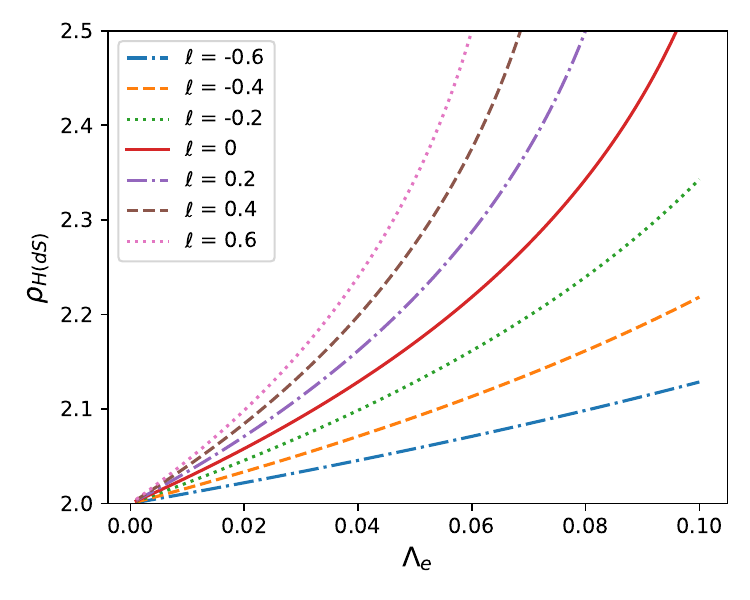}}
	\end{center}
    \caption{Variation of black hole horizon radius with bumblebee parameter $\ell$ (left panel, $\ell = 0$ to $1$, $\Lambda_e = 0.1$) and cosmological constant $\Lambda_e$ (right panel, $\Lambda_e = 0.01$ to $0.2$, $\ell = 0.5$) for Schwarzschild de Sitter-like case ($\Lambda_e > 0$), with $ M = 1 $.}
    \label{dSBHH}
\end{figure}

\begin{figure}[h]
	\begin{center}
        {\includegraphics[width=0.48\textwidth]{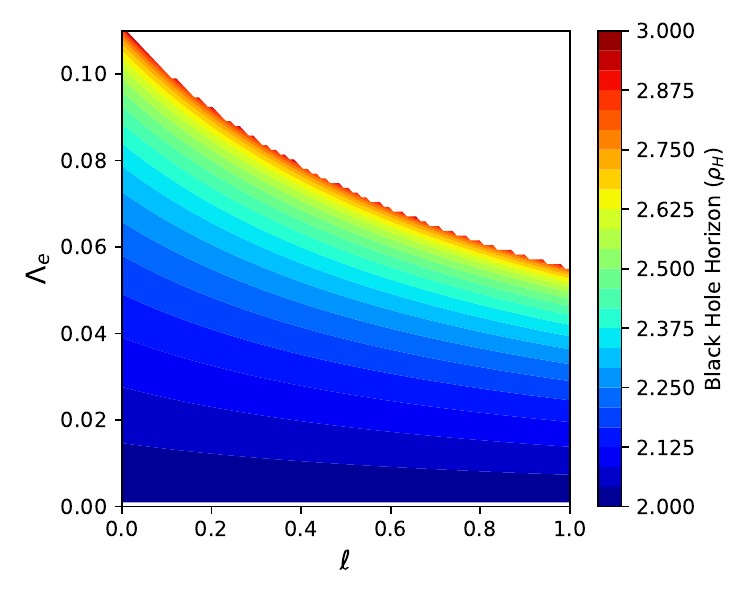}} 
        {\includegraphics[width=0.48\textwidth]{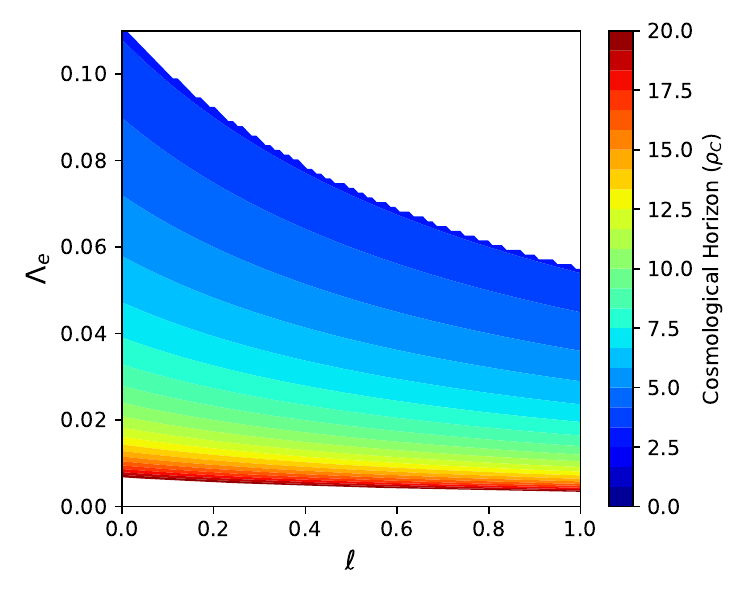}}
	\end{center}
    \caption{Variation of black hole horizon (left panel) and cosmological horizon (right panel) with bumblebee parameter $\ell$ ($\ell = 0$ to $1$) and effective cosmological constant $\Lambda_e$ ($\Lambda_e = 0.01$ to $0.2$) for Schwarzschild de Sitter-like case ($\Lambda_e > 0$), with $ M = 1 $.}
    \label{Dplot}
\end{figure}

\textbf{Case I dS $(\Lambda_{e}>0)$:}  In this case, we find the following two positive real roots,
\begin{equation}
    \rho_{H(dS)} = \frac{2}{\sqrt{\Lambda_{e}(1+\ell)}} \cos \left[ \frac{1}{3}\cos^{-1}(3M\sqrt{\Lambda_{e}(1+\ell)}) + \frac{\pi}{3}\right],
\end{equation}
\begin{equation}
    \rho_{C(dS)} = \frac{2}{\sqrt{\Lambda_{e}(1+\ell)}} \cos \left[ \frac{1}{3}\cos^{-1}(3M\sqrt{\Lambda_{e}(1+\ell)}) - \frac{\pi}{3}\right],
\end{equation}
Here, $\rho_{H}$ and $\rho_{C}$ denote the two true horizons of the spacetime. The larger root $\rho_{C}$ is known as the cosmological horizon, and the smaller root  $\rho_{H}$ is known as the black hole event horizon. Here we consider the dS geometry for which  $\Lambda_{e}>0$ but is very small. Therefore, the horizon radius can be easily obtained in the limit $3M\sqrt{\Lambda_{e}(1+\ell)}<<1$. Now, if we first consider $\cos{\phi} =x$, we have $(\frac{\pi}{2}-\phi)=\sin^{-1}{x}$ for $x<<1$. Therefore, the quantity $\cos^{-1}(3M\sqrt{\Lambda_{e}(1+\ell)})$ in the previous equation can be approximated with $(\frac{\pi}{2} - (3M\sqrt{\Lambda_{e}(1+\ell)})$. Now, the horizon can be given as follows,
\begin{equation}
    \rho_{H} \approx \frac{2}{\sqrt{\Lambda_{e}(1+\ell)}} \cos\left( \frac{\pi}{2} -3M\sqrt{\Lambda_{e}(1+\ell)} \right) = \frac{2}{\sqrt{\Lambda_{e}(1+\ell)}} \sin\left(3M\sqrt{\Lambda_{e}(1+\ell)} \right) = 2 M \left[ 1+ \mathcal{O}(M\sqrt{\Lambda_{e}(1+\ell)})^{2} \right].
\end{equation}
The variation of black hole horizon with bumblebee coupling constant (left panel) and cosmological constant (right panel) for Schwarzschild dS-like case $(\Lambda_{e} >0)$ is illustrated in Fig. \ref{dSBHH}. It is observed that the effective cosmological constant and bumblebee parameter significantly increased the horizon radius. 
\textcolor{black}{In Fig. \ref{Dplot}, the variation of the black hole horizon (left panel) and cosmological horizon (right panel) with the bumblebee parameter and positive cosmological constant for the Schwarzschild-dS-like case is shown. The black hole horizon radii are largest when the bumblebee parameter is smallest and the cosmological constant is highest, while they decrease as the bumblebee parameter increases. Conversely, the cosmological horizon radii decrease with increasing values of both parameters and reach their maximum when the positive cosmological constant is minimal, and the bumblebee parameter is maximal. Physically, the bumblebee parameter alters spacetime curvature, reducing both horizons as it increases, while the cosmological constant drives accelerated expansion, increasing the black hole horizon and reducing the cosmological horizon. These effects highlight the interaction between local spacetime modifications and global dynamics.}


\textbf{Case II AdS $(\Lambda_{e}<0)$:} In this case, there exists only a unique horizon, whose radius is given as,
\begin{equation}
    \rho_{H(AdS)} = -\frac{1}{\mathcal{H}^{1/3}}-\frac{\mathcal{H}^{1/3}}{(1+\ell)\Lambda_{e}}
\end{equation}
where,
\begin{equation}
    \mathcal{H} = (1+\ell)^2 \Lambda_{e}^2 \left( 3M + \sqrt{9M^2-\frac{1}{(1+\ell)\Lambda_{e}}}\right)
\end{equation}
\begin{figure}[h]
	\begin{center}
        {\includegraphics[width=0.48\textwidth]{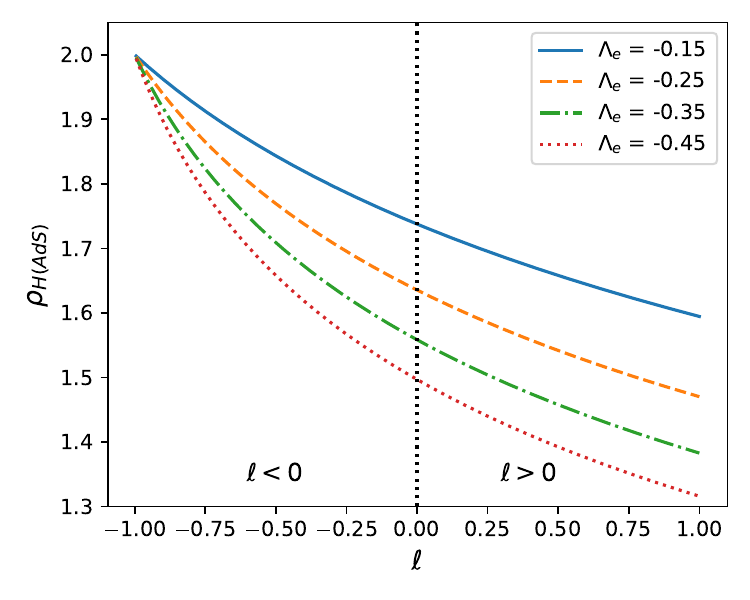}} 
        {\includegraphics[width=0.48\textwidth]{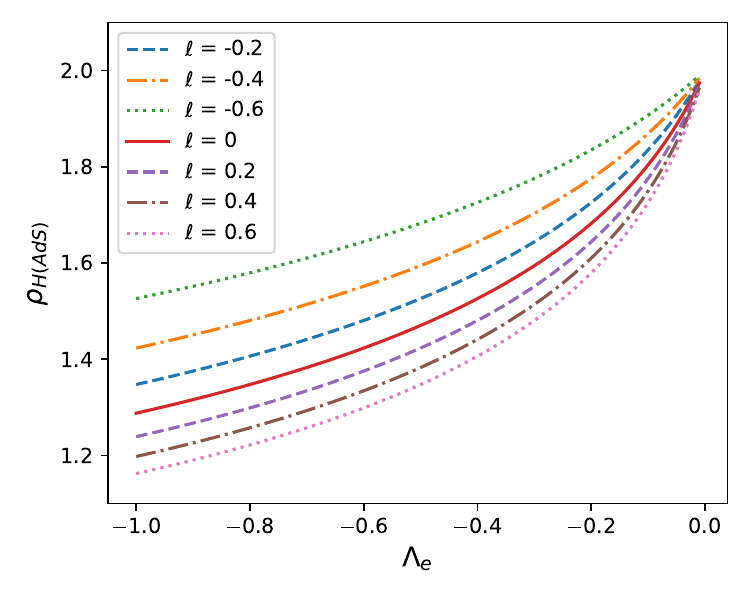}}
	\end{center}
    \caption{Variation of black hole horizon radius with bumblebee parameter $\ell$ (left panel, $\ell = -0.5$ to $0.5$, $\Lambda_e = -0.1$) and cosmological constant $\Lambda_e$ (right panel, $\Lambda_e = -0.2$ to $-0.01$, $\ell = 0.5$) for Schwarzschild anti-de Sitter-like case ($\Lambda_e < 0$), with $ M = 1 $.}
    \label{AdSBHH}
\end{figure} 

The graphical representation of the black hole horizon for the Schwarzschild AdS-like case $(\Lambda_{e} <0)$ is depicted in \figurename{\ref{AdSBHH}}. It can be easily seen that the horizon radius decreased with increasing bumblebee parameter. However, the horizon radius increases with an increase in the effective cosmological constant. It is interesting to observe that for $\ell<0$ the horizon decreased exponentially and became saturated after $\ell>0$.
\section{Thermodynamics} \label{sec3}  

This section analyzes thermal characteristics of Einstein-Bumblebee black holes, incorporating cosmological constant effects. Building upon Kastor's pressure-volume reinterpretation of AdS/dS thermodynamics \cite{Kastor:2009wy}, where pressure $P \equiv -\Lambda/8\pi$ couples to conjugate volume $V$, we establish black hole mass as thermodynamic enthalpy. The revised first law becomes:  

\begin{equation}  
d\mathcal{H} = T dS + V dP,  
\end{equation}  
where enthalpy $\mathcal{H}(S,P)$ replaces internal energy $U(S,V)$. Thermodynamic volume emerges through the derivative:  
\begin{equation}  
V = \left( \frac{\partial \mathcal{H}}{\partial P} \right)_{S}.  
\end{equation}  

Considering only black hole horizons ($\Lambda_e<0$) ensures thermal equilibrium, as distinct horizon temperatures would induce thermal fluxes. The mass-horizon relation and surface gravity expressions are:  
\begin{equation}  
M = \frac{\rho_H}{2}\left( 1-\frac{\Lambda_e(1+\ell)\rho_H^{2}}{3}\right), \quad \mathcal{K} = \frac{1}{2} \left( \frac{2M}{\rho_H^{2}} - (1-\ell) \frac{2\Lambda_e}{3}\rho_H\right).  
\end{equation}  

The Hawking temperature derived from surface gravity,  
\begin{equation}  
T_{\mathcal{K}} = \frac{1}{4\pi} \left( \frac{2M}{\rho_H^{2}} - (1-\ell) \frac{2\Lambda_e}{3}\rho_H\right),  
\end{equation}  
matches the thermodynamic temperature obtained via enthalpy differentiation:  
\begin{equation}  
T_H = \left(\frac{\partial \mathcal{H}}{\partial S}\right)_{P} = \frac{1-\Lambda_e(1+\ell)\rho_H^{2}}{4\pi\rho_H}.  
\end{equation}  

\begin{figure}[htbp]  
	\centering  
    \includegraphics[width=0.48\textwidth]{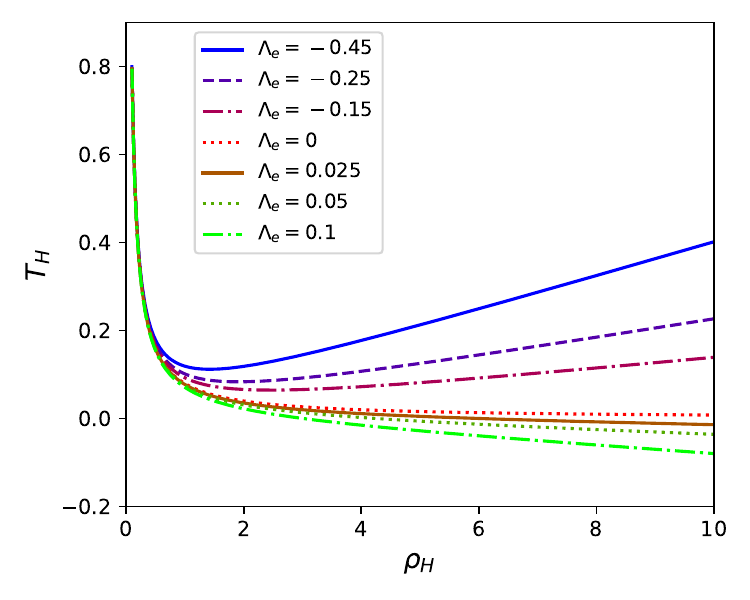}  
    \includegraphics[width=0.48\textwidth]{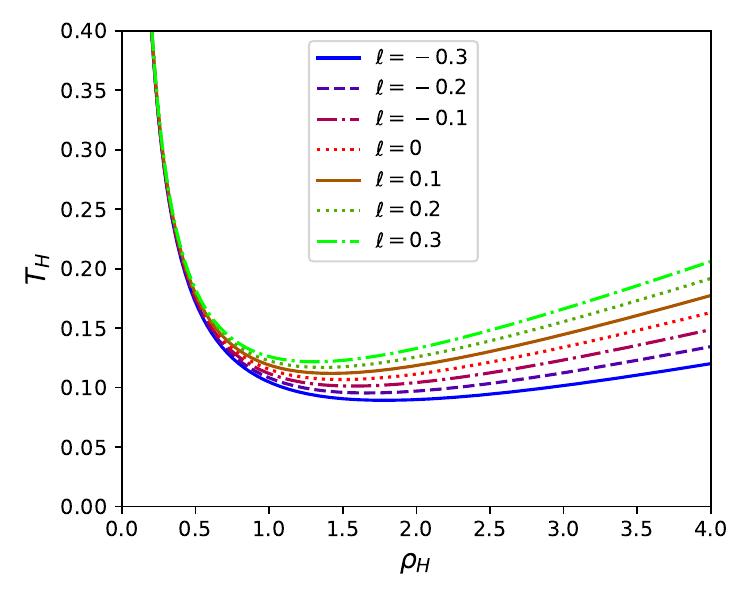}  
	\caption{Thermal emission characteristics versus event horizon radius: (Left) Varying effective cosmological constant $\Lambda_e$ ($\Lambda_e = -0.1$ to $-0.01$, $\ell = 0.05$). (Right) Varying bumblebee parameter $\ell$ ($\ell = -0.1$ to $0.1$, $\Lambda_e = -0.045$), with $ M = 1 $.}
    \label{HawkingT}  
\end{figure}  

Nonlinear bumblebee field couplings necessitate modifying the first law through energy-momentum tensor corrections:  
\begin{equation}  
d \mathcal{M} = T dS + V dP, \quad \mathcal{C} = 1- \frac{3(1+\ell)-(1+\ell)\Lambda_e\rho_H^{2}}{3\rho_H^{2}},  
\end{equation}  
where $\mathcal{C}$ compensates for $T_0^0$ tensor contributions (detailed in \cite{wang2024thermodynamics}).  

Entropy and volume relations maintain standard forms despite modifications:  
\begin{equation}  
S = \pi \rho_H^2, \quad V = \frac{4\pi\rho_H^3}{3}.  
\end{equation}  

The equation of state, expressed through specific volume $v \equiv 2\rho_H$, becomes:  
\begin{equation}  
P = \frac{T}{v(1+\ell)}-\frac{1}{2 \pi v^2 (1+\ell)}.  
\end{equation}  

Critical parameters derived from inflection conditions $\partial_v P = \partial_v^2 P = 0$ yield:  
\begin{equation}  
P_c = \frac{1.5}{\pi^3} \left( 1-\frac{1}{2 \pi (1+\ell)} \right), \quad v_c = \frac{\pi^2}{(1+\ell)}\sqrt{\frac{2}{3}}, \quad T_c = \frac{1}{\pi}\sqrt{\frac{3}{2}}.  
\end{equation}  

The critical ratio $P_c v_c/T_c \approx 0.442$ exceeds van der Waals predictions (0.375), aligning with charged AdS black hole behavior. Notably, $T_c$ remains $\ell$-independent while $v_c$ decreases with increasing Lorentz violation - a distinctive feature absent in standard GR solutions.  

\begin{figure}[htbp]  
	\centering  
    \includegraphics[width=0.48\textwidth]{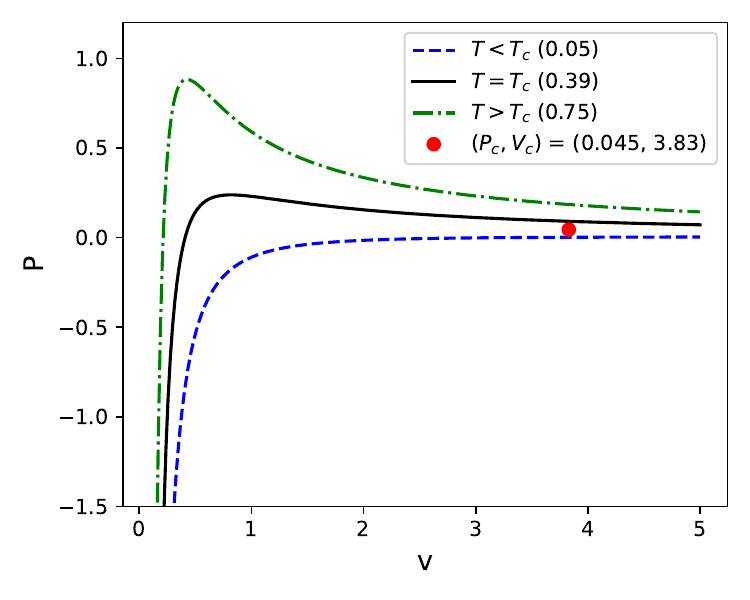}  
    \includegraphics[width=0.48\textwidth]{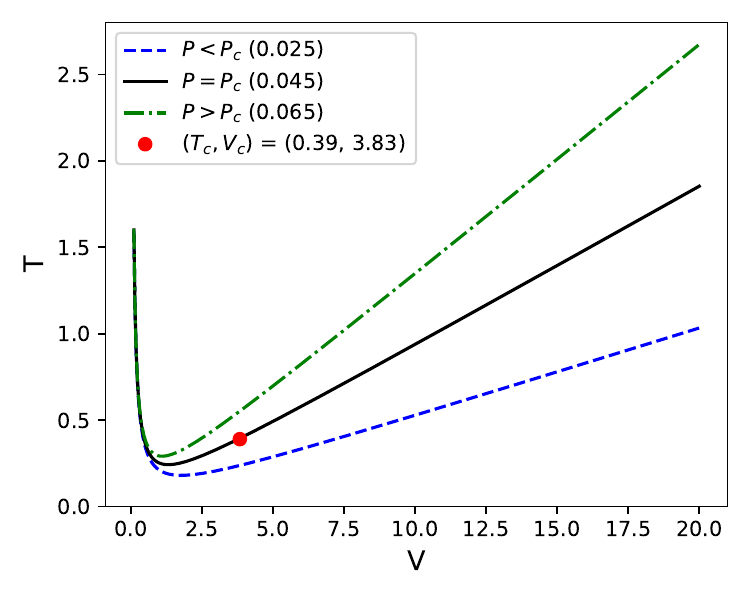}  
	\caption{Phase transition signatures: (Left) $ P-v $ isotherms near criticality ($\ell = 0.1$, $\Lambda_e = -0.1$, $ P = 0.01$ to $0.1$). (Right) $ T-v $ isobars ($\ell = 0.1$, $\Lambda_e = -0.1$, $ T = 0.1$ to $0.5$), with $ M = 1 $. Dashed lines indicate unstable phases.}
    \label{PVT}  
\end{figure}  

Figure \ref{HawkingT} demonstrates thermal profile modifications: increasing $\Lambda$ suppresses temperatures at fixed $\rho_H$, while positive $\ell$ enhances emission rates. The $\ell<0$ regime exhibits suppressed thermal output, suggesting observational constraints on Lorentz violation through accretion disk spectroscopy.

\section{Shadow analysis} \label{sec4}
To determine the black hole shadow for the Schwarzschild metric, it's essential to study the behavior of photons (light rays) around the black hole. Specifically, we focus on photon orbits, since the shadow is formed by the photon paths that either orbit near the black hole or are captured by it, never reaching the distant observer. A straightforward approach to this analysis is outlined in Ref.  
 \cite{Perlick:2021aok,Vagnozzi:2022moj}, which we will follow here. \textcolor{black}{Furthermore, despite the known rotational nature of astrophysical black holes such as M87* and Sgr A*, we adopt a spherically symmetric black hole model for shadow analysis. The main reason for this choice is motivated by previous works in Ref. \cite{Vagnozzi:2022moj} that demonstrate the small effect of spin on the shadow size, typically modifying the radius by less than 12\% for moderate spin values. Moreover, current observational constraints on Sgr A*'s spin remains inconclusive, with some recent dynamical estimates suggesting a low spin ($a* \lesssim 0.1$). Given these factors, neglecting spin simplifies the analysis while still capturing the essential physics of black hole shadows.}

We review the calculation of the shadow radius in spherically symmetric space-times, where the gravitationally lensed image of a photon sphere, if present, forms the black hole shadow for an observer at infinity. Although a photon sphere often leads to a shadow, it is not strictly required for a space-time to cast one (see Ref.~\cite{Joshi:2020tlq}). For a static, spherically symmetric, asymptotically flat space-time with a time-like Killing vector and diagonal metric, the line element can be expressed as
\begin{eqnarray}
{\rm d}s^2 = -A(\rho){\rm d}t^2 + B(\rho){\rm d}\rho^2+C(\rho){\rm d}\Omega^2.
\label{eqstaticsphericallysymmetricasymptoticallyflat}
\end{eqnarray}

Here, $d\Omega$  represents the differential unit of solid angle. We define $A(\rho)=-g_{tt}(\rho)$ as the “metric function” and, for convenience, $C(\rho)=g_{\theta\theta}=g_{\phi\phi}/\sin^2\theta$ as the “angular metric function,” though this latter term is not widely used (\ref{eqstaticsphericallysymmetricasymptoticallyflat}). In asymptotically flat space-times, $A(\rho) \to 1$ as $\rho \to \infty$, we will maintain a more general treatment for broader applicability.

We now define the function $h(\rho)$ 
 (as introduced in, for example, Ref.~\cite{Perlick:2021aok}):

\begin{eqnarray}
h(\rho) \equiv \sqrt{\frac{C(\rho)}{A(\rho)}}\,.
\label{eq:hr}
\end{eqnarray}

To find the radial coordinate of the photon sphere, $\rho_{\rm ph}$, we solve the following implicit equation for the metric in Eq.~(\ref{eqstaticsphericallysymmetricasymptoticallyflat}).

\begin{eqnarray}
\frac{\rm d}{{\rm d}\rho} \left [ h^2(\rho_{\rm ph}) \right ] =0.
\label{eq:photonspherehr}
\end{eqnarray}
We can rearrange this equation into the following form:
\begin{eqnarray}
C'(\rho_{\rm ph})A(\rho_{\rm ph})-C(\rho_{\rm ph})A'(\rho_{\rm ph})=0\,.
\label{eq:photonspherecomplete}
\end{eqnarray}

Note that the prime represents differentiation with respect to $\rho$.  If one takes $C(\rho)=\rho^2$, Eq.~(\ref{eq:photonspherecomplete}) simplifies to the well-known expression:

\begin{eqnarray}
A(\rho_{\rm ph})-\frac{1}{2}\rho_{\rm ph}A'(\rho_{\rm ph})=0\,.
\label{eq:photonsphere}
\end{eqnarray}

The gravitational lensing of the surface at $\rho_{\rm ph}$ produces the shadow radius $r_{\rm sh}$, which is given by (see, for example, Refs.\cite{Psaltis:2007rv,Cunha:2018acu,Dokuchaev:2019jqq,EventHorizonTelescope:2020qrl,Perlick:2021aok}):
\begin{eqnarray}
r_{\rm sh} = \sqrt{\frac{C(\rho)}{A(\rho)}}\Bigg\vert_{\rho_{\rm ph}}\,.
\label{eq:rshcomplete}
\end{eqnarray}
With $C(\rho)=\rho^2$ Eq.~(\ref{eq:rshcomplete}) becomes the shadow radius of the black hole:
\begin{eqnarray}
r_{\rm sh} = \frac{\rho}{\sqrt{A(\rho)}}\Bigg\vert_{\rho_{\rm ph}}\,.
\label{eq:rsh}
\end{eqnarray} 
\begin{figure*}
    \centering
    {\includegraphics[width=0.48\textwidth]{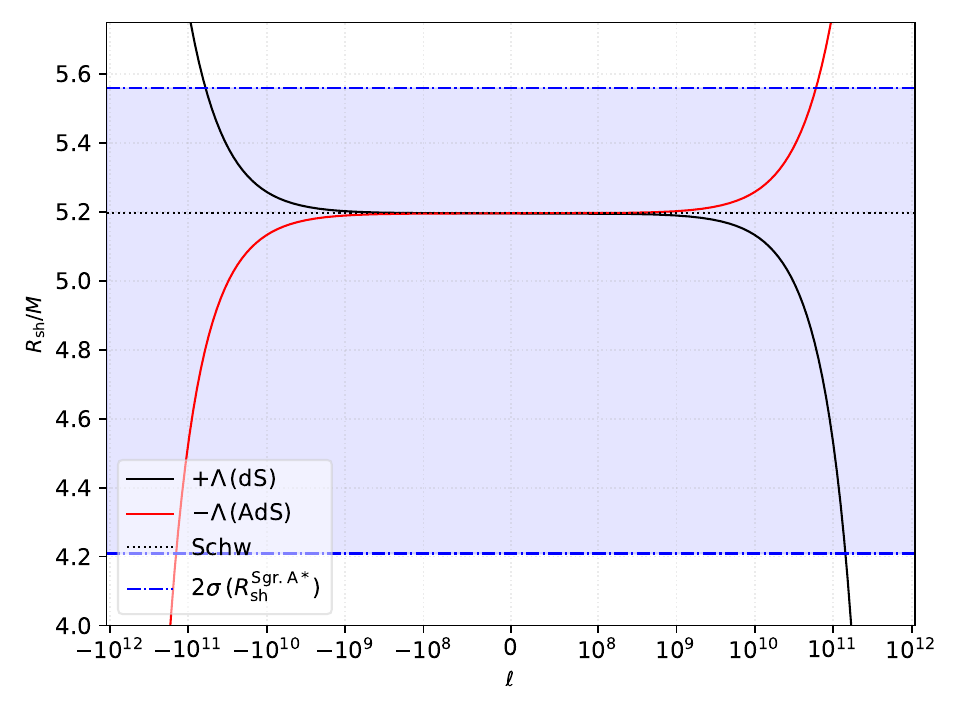}} 
    {\includegraphics[width=0.48\textwidth]{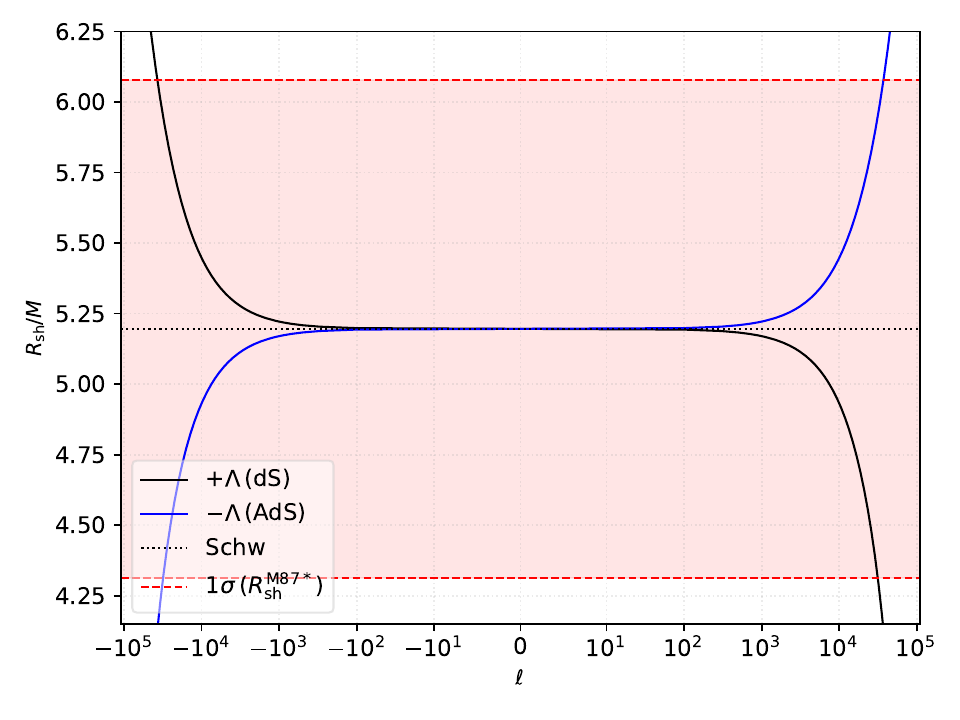}}
    \caption{Constraints on the bumblebee parameter $\ell$ ($\ell = -10^{11}$ to $10^{11}$ for Sgr A*, $-34311$ to $34309$ for M87*). Left: $\Lambda_e M^2 = \pm 4.46 \times 10^{-33}$. Right: $\Lambda_e M^2 = \pm 1.02 \times 10^{-26}$, with $ M = 1 $.}
    \label{shadow_cons}
\end{figure*}
For the Schwarzschild black hole, the shadow radius can be obtained as $r_{\rm sh}=3\sqrt{3}M$. This outcome is derived from the geometry of photon orbits near the black hole, with the shadow radius directly linked to the photon sphere. Furthermore, due to the high degree of spherical symmetry inherent in the Schwarzschild solution, the resulting shadow remains a perfect circle with radius $r_{\rm sh}$ on the image plane of a distant observer. This symmetry ensures that the shadow's shape and size are unaffected by the observer's inclination angle, making it independent of the viewing direction.

Due to the proximity of Sgr A*, the focus is placed on a static observer located at a distance $\rho_O$. In this context, the angular size of the black hole shadow, $\alpha_{\rm sh}$, is given by (see, e.g., Ref.~\cite{Perlick:2021aok}):

\begin{eqnarray}
\sin^2\alpha_{\rm sh} = \frac{\rho^2_{\rm ph}}{A(\rho_{\rm ph})}\frac{A(\rho_O)}{\rho_O}\,.
\label{eq:alphashnotasymptoticallyflat}
\end{eqnarray}
In the physically relevant small-angle approximation, it is straightforward to observe that the shadow size is determined by:
\begin{eqnarray}
r_{\rm sh} = \rho_{\rm ph}\sqrt{\frac{A(\rho_O)}{A(\rho_{\rm ph})}} \sim 3\sqrt{3M} - \frac{\sqrt{3}\rho_O^2(1+\ell)\Lambda_e}{2} + \mathcal{O}(M^2)\,.
\label{eq:rshnotasymptoticallyflat}
\end{eqnarray}

Eq.(\ref{eq:rshnotasymptoticallyflat}) clearly shows the explicit dependence of the shadow size on the observer's position. For an observer situated far from a black hole described by an asymptotically flat metric, Eq.(\ref{eq:rshnotasymptoticallyflat}) easily reduces to Eq.~(\ref{eq:rsh}). This occurs because, at large distances, the metric function approaches $A(\rho_O) \approx 1$, as the gravitational influence of the black hole becomes negligible. Thus, when the observer is sufficiently distant, the shadow size naturally simplifies to the familiar expression.

The results from the EHT collaboration \cite{EventHorizonTelescope:2019dse,EventHorizonTelescope:2022wkp,EventHorizonTelescope:2021dqv,Vagnozzi:2022moj} are used to find the constraint in $\ell$ using the shadow radius of Sgr A* \textcolor{black}{($ 4.209M \leq R_{\rm Sch} \leq 5.560M $ at $2\sigma$ level) and M87* ($ 4.313M \leq R_{\rm Sch} \leq 6.079M $ at the $1\sigma$ level)}. These results are presented in Fig. \ref{shadow_cons}. As noticed, there is a huge range for $\ell$ when we constrain it using the Sgr A*. That is, $-10^{11} \lesssim \ell \lesssim 10^{11}$. Using the parameters for M87*, we find smaller values for the parameter: $-34311 \lesssim \ell \lesssim 34309$. Analytically, these values can be obtained through
\begin{equation}
    \ell = -\left(1 + \frac{2\sqrt{3}\delta}{3 \rho_O^2 M \Lambda_e} \right),
\end{equation}
where $\delta$ is the difference between $r_{\rm sh}$ and both the upper and lower bounds due to uncertainties. Furthermore, we see the importance and the dependence of $\ell$ on the position of the observer $\rho_O$ relative to the black hole.

\section{Analysis of the deflection angle in the weak field limit} \label{sec5}
In this section, we study the weak deflection angle of a black hole in Einstein-bumblebee gravity using the generalized GW method. To investigate the null geodesics, we introduce the optical metric constrained to the equatorial plane $(\theta = \pi/2)$, which simplifies the analysis and can be expressed as follows \cite{Huang:2023bto},
\begin{equation}
    dl^2 = \alpha_{\rho\rho}(\rho) d\rho^{2} + \alpha_{\phi\phi}(\rho) d\phi^{2}.
\end{equation}
The Gaussian curvature corresponding to Riemannian component ($M^{\alpha2}$) can be calculated as \cite{Werner:2012rc},
\begin{equation}
    \int \mathcal{K} \sqrt{\alpha} d\rho = \int -\frac{\partial}{\partial \rho} \left( \frac{\sqrt{\alpha}}{\alpha_{\rho\rho}} \Gamma^{\phi}_{r\phi}\right) = - \frac{\alpha_{\phi\phi},\rho}{2\sqrt{\alpha}}. 
\end{equation} 
Consequently, the surface integral of the Gaussian curvature can be expressed as,
\begin{equation}
    \int \int_{D_{\infty}} \mathcal{K} ds = \int_{\phi_{\rm S}}^{\phi_{\rm R}} \int_{\rho_{\gamma}}^{\infty} \mathcal{K} \sqrt{\alpha} dr d\phi .
\end{equation}
Here, $\rho_{\gamma}$ is the radial curvature of geodesics who along the source and observer, $\phi_{\rm S}$ and $\phi_{\rm R}$ are the azimuthal coordinates of the source and receiver, respectively. $\alpha$ represents the metric determinant in the optical framework. The above expression can be expressed as,
\begin{equation}
    \int \int_{D_{\infty}} \mathcal{K} ds = \int_{\phi_{\rm S}}^{\phi_{\rm R}} [H(\infty) - H (\rho_{\gamma})] d\phi,
\end{equation}
here, 
\begin{equation} \label{Hvalue}
    H(\rho) = - \frac{\alpha_{\phi\phi},\rho}{2\sqrt{\alpha}} .
\end{equation}
On accounts of the asymptotics of $M^{\alpha2}$, we have $H(\infty) = -1$ and the geometric expression of finite-distance deflection angle reads as \cite{Huang:2022iwl},
\begin{equation}
    \hat{\alpha} = \int_{\phi_{\rm S}}^{\phi_{\rm R}} [1 + H (\rho_{\gamma})] d\phi + \int_{\gamma} k dl.
\end{equation}
Here, k is the geodesic curvature of $\gamma$ in  $M^{\alpha2}$ and can be evaluated using the following expression \cite{PhysRevD.96.104037},
\begin{equation}
    k = \frac{\beta_{\phi, \rho}}{\sqrt{\alpha \alpha^{\theta\theta}}}
\end{equation}
For stationary BHs $\int_{\gamma} k dl=0$, hence the expression for finite-distance deflection angle reads,
\begin{equation} \label{finaldefformula}
    \hat{\alpha} = \int_{\phi_{\rm S}}^{\phi_{\rm R}} [1 + H (\rho_{\gamma})] d\phi. 
\end{equation}
Our analysis fuscous on the trajectory of massless particles confined to the equatorial plane, and the corresponding orbit equation can be derived as follows,
\begin{equation}\label{eorb}
    \left( \frac{du}{d\phi} \right)^{2} = \frac{1}{(1+\ell)b^{2}} - \frac{u^{2}}{(1+\ell)} + \frac{2Mu^{3}}{(1+\ell)} + \frac{(1+\ell)\Lambda_{e}}{3},
\end{equation}
here, $u=1/\rho$. Thus, we get the perturbation solution of the orbit equation as given below,
\begin{equation} \label{trajectory}
    u = \frac{\sin\frac{\phi}{\sqrt{(1+\ell)}}}{b} + \frac{M(1+ \cos^{2}\frac{\phi}{\sqrt{(1+\ell)}})}{b^{2}}  + \frac{b\Lambda_{e}\sin\frac{\phi}{\sqrt{(1+\ell)}}}{6} + \frac{M\Lambda_{e}(1+ \cos^{2}\frac{\phi}{\sqrt{(1+\ell)}})}{3}  + \mathcal{O} (M^{2}, \Lambda_{e}^{2}),
\end{equation}
which we can determine the solution using an iteration method in terms of $\phi$,
\begin{equation} \label{ISlimit}
    \phi(u) = 
\begin{cases} 
\Phi(u) & \text{if } |\phi| < \frac{\pi}{2}, \\
\pi\sqrt{(1+\ell)} - \Phi(u) & \text{if } |\phi| \geq \frac{\pi}{2},
\end{cases}
\end{equation}
where,
\begin{equation}
\begin{split}
    &\Phi(u) = \sqrt{(1+\ell)}\arcsin(bu) + M \sqrt{(1+\ell)} \frac{b^2 u^2-2}{b \sqrt{1 - b^2 u^2}} - \frac{ (1+\ell)\sqrt{(1+\ell)} \Lambda_{e} b }{6 u \sqrt{1 - b^2 u^2}} + \\
    &M \Lambda_{e} \sqrt{(1+\ell)} \frac{b (-3b^4 u^4 + 8b^2 u^2 - 4)}{6 (1 - b^2 u^2)^{3/2} } + \mathcal{O} (M^2, \Lambda^2).
    \end{split}
\end{equation}

\begin{figure}[htbp] 
	\begin{center}
        {\includegraphics[width=0.48\textwidth]{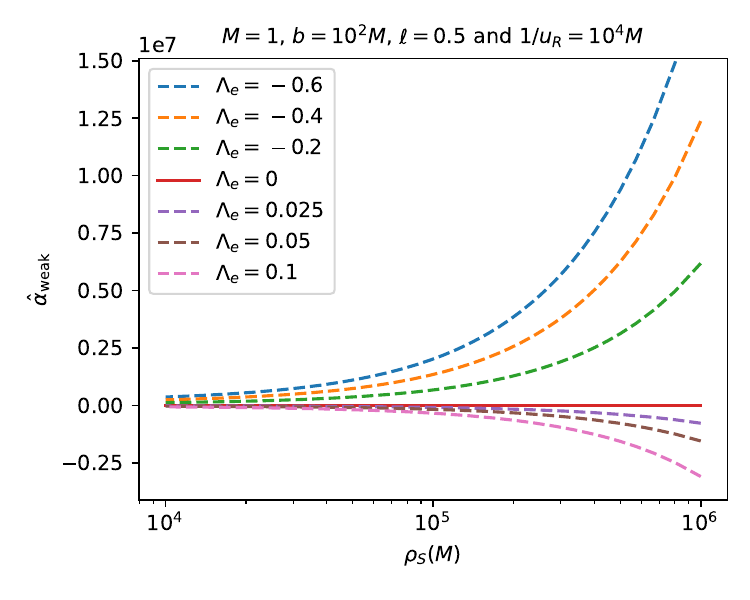}} 
        {\includegraphics[width=0.48\textwidth]{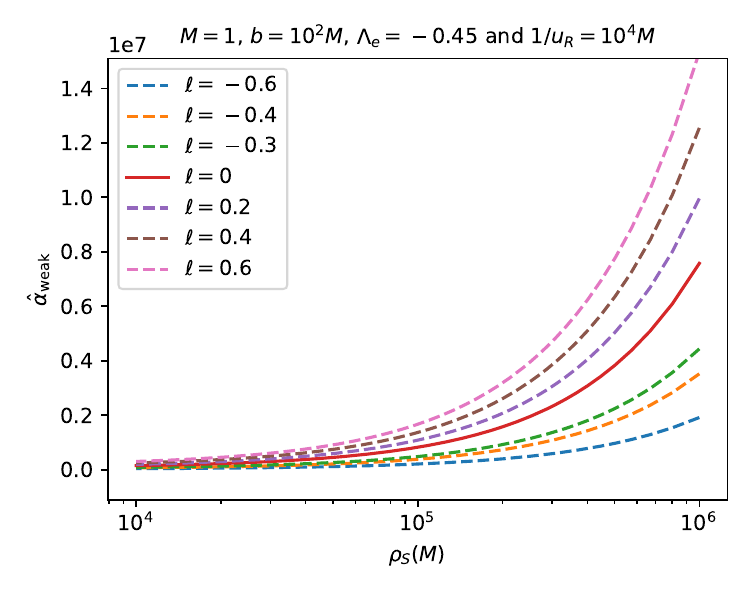}}
	\end{center}
    \caption{Variation of weak-field deflection angle with source distance: (Left) Varying cosmological constant $\Lambda_e$ ($\Lambda_e = -0.1$ to $0.1$, $\ell = 0.5$). (Right) Varying bumblebee parameter $\ell$ ($\ell = 0.1$ to $1$, $\Lambda_e = -0.1$), with $ M = 1 $, $ b = 10^2 M $, and $\rho_R = 10^4 M $.}
    \label{deflectionangle}
\end{figure}

Now, the optical metric of the black hole in Einstein-bumblebee gravity corresponding to $M^{\alpha2}$ is given as,
\begin{equation} \label{Phivalue}
    dl^{2} = \frac{(1+\ell)d\rho^{2}}{(1-\frac{2M}{\rho}-(1-l)\frac{\Lambda_{e}}{3}\rho^{2})^2} + \frac{\rho^{2}d\phi}{1-\frac{2M}{\rho}-(1-l)\frac{\Lambda_{e}}{3}\rho^{2}}.
\end{equation}
Substituting Eq. \eqref{Phivalue} into Eq. \eqref{Hvalue} leads to,
\begin{equation}
    H (\rho) = -\frac{1}{\sqrt{1+\ell}} + \frac{2M}{\rho\sqrt{1+\ell}} -\frac{(1+\ell)\Lambda_{e}\rho^2}{6\sqrt{1+\ell}} + \frac{3M^2}{\rho^2 \sqrt{1+\ell}} + \frac{(1+\ell)M\Lambda_{e}\rho^2}{2\sqrt{1+\ell}}.
\end{equation}
Now, from Eq. \eqref{ISlimit}, we have $\phi_{\rm S} =\Phi(u_{S})$ and $\phi_{\rm R} =\pi\sqrt{(1+\ell)} -\Phi(u_{R})$. Therefore, using Eq. \eqref{finaldefformula} the deflection angle in the weak-field regime is obtained as follows,
\begin{equation} \label{weakdl}
    \begin{split}
    & \hat{\alpha}_{\text{weak}}  = (\sqrt{1 + \ell} - 1) \left[ \pi - \arcsin(bu_R) - \arcsin(bu_S) \right] 
    + \frac{M}{b} \left[ \frac{2\sqrt{1+\ell}-b^2u_{R}^2(1+\sqrt{1 + \ell})}{\sqrt{1-b^2u_{R}^2}} + \frac{2\sqrt{1+\ell}-b^2u_{S}^2(1+\sqrt{1 + \ell})}{\sqrt{1-b^2u_{S}^2}} \right]   \\
    & - \frac{\Lambda_{e}b(1+\ell)\sqrt{1+\ell}}{6} \left[ \frac{\sqrt{1-b^2u_{R}^2}}{u_{R}} + \frac{\sqrt{1-b^2u_{S}^2}}{u_{S}} \right] + \frac{M\Lambda_{e}b(1+\ell)\sqrt{1+\ell}}{6} \left[ \frac{1}{\sqrt{1-b^2u_{R}^2}} + \frac{1}{\sqrt{1-b^2u_{S}^2}}\right]
    + \mathcal{O}(M^2, \Lambda_{e}^2).
\end{split}
\end{equation}
Here, we have used the trajectory Eq. \eqref{trajectory} to obtain the weak-field deflection angle. The result obtained in Eq. \eqref{weakdl} represents the weak-field deflection angle of the black hole in bumblebee gravity. Without the presence of a cosmological constant, the weak-field deflection angle simplifies to that of a Schwarzschild-like black hole within the framework of the bumblebee gravity model.  When $\rho_{\rm S} \xrightarrow{} b$ and $\rho_{\rm R} \xrightarrow{} b$, i.e., the source or the receiver approaches the least impact radius, the gravitational deflection angle will diverge. 
In Fig. \ref{deflectionangle}, we illustrate the finite-deflection angle with source distance for distinct values of cosmological constant and bumblebee parameter. Here, we fixed $M=1$, $b=10^2M$, and $\rho_{\rm R}=10^4M$ and obtained the bending angle on the log scale. For a fixed parameter $\ell=0.5$, it is observed that the deflection angle increases as the distance to the source increases for negative values of the cosmological constant, and the effect is more pronounced for smaller (more negative) $\Lambda_{e}$. Specifically, the weak-field deflection angle is stronger for lesser values of $\Lambda_{e}$. Additionally, for a fixed $\Lambda_{e}$, the deflection angle increases with the source distance. When comparing different values of $\ell$, it was noted that the deflection angle is significantly larger for $\ell<1$ compared to $\ell>1$. Furthermore, the deflection angle reaches its minimum as the bumblebee parameter approaches unity.

\textcolor{black}{It is equally essential to compare the obtained results with those of the standard Schwarzschild case for a comprehensive analysis. For instance, our results indicate that the weak deflection angle in Einstein-Bumblebee gravity exhibits deviations from the standard Schwarzschild case, where the leading-order term scales as $4M/b$. In particular, for $\ell > 0$, the deflection angle is enhanced, while for $\ell < 0$, it is reduced. Additionally, the incorporation of the cosmological constant leads to further modifications: for negative $\Lambda$, the deflection angle increases compared to the Schwarzschild scenario, whereas for positive $\Lambda$, it decreases. These effects become more pronounced for larger impact parameters, suggesting that precise gravitational lensing observations could serve as potential probes of Lorentz symmetry violation.}

\section{Strong-Field Light Bending Analysis} \label{sec6}  
This section quantifies light deflection extremes in the strong gravitational regime of Einstein-Bumblebee black holes. Building on Tsukamoto's analytical framework \cite{tsukamoto2017deflection}, we extend the methodology to spacetime geometries lacking asymptotic flatness. The orbital trajectory equation (Eq.~\eqref{eorb}) is reformulated as:  

\begin{equation}  
\label{eq.43}  
\left( \frac{d\rho}{d\phi} \right)^2 = \frac{R(\rho) \rho^2}{B(\rho)},  
\end{equation}  
where the radial function  
\begin{equation}  
\label{eq.44}  
R(\rho) = \frac{A(\rho_0) \rho^2}{A(\rho) \rho_0^2} - 1  
\end{equation}  
incorporates the metric component $A(\rho)$ evaluated at the closest approach $\rho_0$ (Eq.~\eqref{coeff}). Solving Eq.~\eqref{eq.43} produces the deflection angle expression \cite{tsukamoto2017deflection, bozza2002gravitational}:  
\begin{equation}  
\begin{split}  
\label{eq.45}  
\alpha(\rho_0) &= I(\rho_0) - \pi \\  
&= 2 \int_{\rho_0}^{\infty} \frac{d\rho}{\sqrt{\frac{R(\rho) C(\rho)}{B(\rho)}}} - \pi.  
\end{split}  
\end{equation}  

To resolve the integral's divergent behavior near $\rho_0$, we employ a series expansion about $\rho = \rho_0$, separating the integral into regular ($\kappa_R$) and divergent ($\kappa_D$) components. Introducing the substitution $z \equiv 1 - \frac{\rho_0}{\rho}$ restructures Eq.~\eqref{eq.45} as:  
\begin{equation}  
I(\rho_0) = \int_0^1 \left[ \kappa_D(z, \rho_0) + \kappa_R(z, \rho_0) \right] dz,  
\end{equation}  
where $\kappa_D$ and $\kappa_R$ respectively govern logarithmic divergence and finite contributions (detailed in \cite{tsukamoto2017deflection, bozza2002gravitational}). The resultant SDA becomes:  
\begin{equation}  
\label{eq.48}  
\hat{\alpha}_{\text{str}} = -\bar{a} \log \left( \frac{b_0}{b_{\text{crit}}} - 1 \right) + \bar{b} + O\left( \frac{b_0}{b_{\text{crit}}} - 1 \right) \log \left( \frac{b_0}{b_{\text{crit}}} - 1 \right),  
\end{equation}  
with coefficients $\bar{a}$, $\bar{b}$ encoding spacetime curvature effects:  
\begin{equation}  
\label{eq.49}  
\bar{a} = \sqrt{\frac{2 B(\rho_{\text{ps}}) A(\rho_{\text{ps}})}{2 A(\rho_{\text{ps}}) - A''(\rho_{\text{ps}}) \rho_{\text{ps}}^2}},  
\end{equation}  
\begin{equation}  
\label{eq.50}  
\bar{b} = \bar{a} \log \left[ \rho_{\text{ps}} \left( \frac{2}{\rho_{\text{ps}}^2} - \frac{A''(\rho_{\text{ps}})}{A(\rho_{\text{ps}})} \right) \right] + I_R(\rho_{\text{ps}}) - \pi.  
\end{equation}  
Here, $\rho_{\text{ps}}$ denotes photon sphere radius, and $I_R$ represents the regularized integral:  
\begin{equation}  
\label{eq.53}  
I_R(\rho_0) \equiv \int_0^1 \left[ f_R(z, \rho_0) - f_D(z, \rho_0) \right] dz,  
\end{equation}  
with trajectory functions  
\begin{equation}  
\label{eq.54}  
f_R(z, \rho_0) = \frac{2 \rho_0}{\sqrt{G(z, \rho_0)}}, \quad G(z, \rho_0) = R(\rho) C(\rho) A(\rho) (1 - z)^4.  
\end{equation}  
At critical approach ($\rho_0 = \rho_{\text{ps}}$), these reduce to:  
\begin{equation}  
\label{eq.55}  
f_R(z, \rho_{\text{ps}}) = \frac{2 \rho_{\text{ps}}}{\sqrt{\sum_{m=2} c_m(\rho_{\text{ps}}) z^m}}, \quad f_D(z, \rho_{\text{ps}}) = \frac{2 \rho_{\text{ps}}}{\sqrt{c_2 z^2}}.  
\end{equation}  
Evaluating $I_R(\rho_{\text{ps}})$ yields:  
\begin{equation}  
I_R(\rho_{\text{ps}}) = \log \left( \frac{144}{(1 + \sqrt{3})^{1/4}} \right).  
\end{equation}  

The final deflection angle expression incorporates Lorentz-violating ($\ell$) and cosmological ($\Lambda_e$) parameters:  
\begin{equation}  
\hat{\alpha}_{\text{str}} = -\log \left( \frac{b_0}{b_{\text{crit}}} - 1 \right) + \log \left[ \frac{-6}{-1 + 9 \Lambda_e (\ell + 1) M^2} \times \frac{144}{(1 + \sqrt{3})^{1/4}} \right] - \pi + O\left( \frac{b_0}{b_{\text{crit}}} - 1 \right) \log \left( \frac{b_0}{b_{\text{crit}}} - 1 \right).  
\end{equation}  

Impact parameter calculations via Eq.~\eqref{eorb} reveal:  
\begin{equation}  
b_0^2 = \frac{-6 \rho_0^3}{2 \Lambda (\ell + 1) \rho_0^3 - 3 \rho_0 + 3 M},  
\end{equation}  
asymptotically approaching $b_{\text{crit}}$ for $\rho_0 \to 3M$.  

\begin{figure}[htbp]  
    \centering  
    \includegraphics[width=0.48\textwidth]{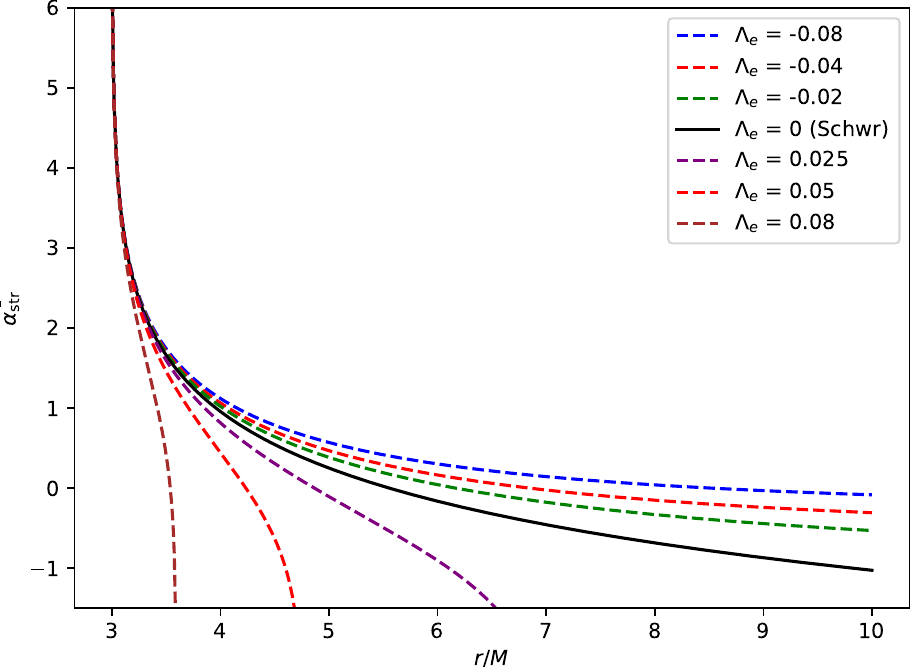}  
    \includegraphics[width=0.48\textwidth]{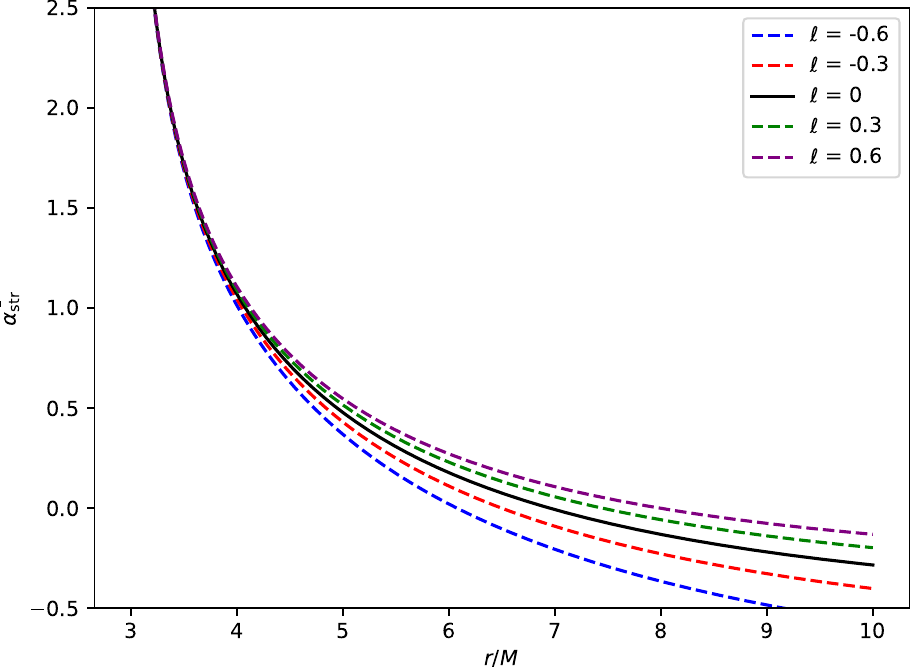}  
    \caption{Dependence of strong deflection angle on closest approach distance $\rho_0$: (Left) Varying effective cosmological constant $\Lambda_e$ ($\Lambda_e = -0.1$ to $0.1$, $\ell = 0.05$). (Right) Varying bumblebee parameter $\ell$ ($\ell = -0.1$ to $0.1$, $\Lambda_e = -0.045$), with $ M = 1 $, $\rho_0 = 3M$ to $10M $.}
    \label{deflectionangle2}  
\end{figure}  

Key modifications from GR emerge through $\ell$ and $\Lambda$ dependencies in both logarithmic coefficients and critical impact parameters, providing observational signatures testable through high-precision lensing measurements.

In the strong field limit, we observe a notable dependence of the deflection angle on the bumblebee parameter, $\ell$. For $\ell>0$, the deflection angle increases proportionally with increasing $\ell$. Conversely, for $\ell<0$, the deflection angle decreases at a similar rate. For a constant value of $\ell$, a highly negative cosmological constant leads to an increase in the deflection angle. In contrast, when the cosmological constant becomes highly positive, we see a significant decrease in the deflection angle. Notably, the effects of a highly positive cosmological constant exhibit considerable deviation from those observed with negative values.

Observations obtained from the EHT have shown that strong gravitational lensing produces a bright photon ring surrounding the central shadow of a black hole \cite{EventHorizonTelescope:2019dse}. For a non-rotating black hole ($a_{*} = 0$), the angular radius of the photon ring is given by:

\begin{equation}\label{eq65}
\theta_{\rm p} = \frac{\sqrt{27}GM}{c^{2}D} = 18.8\left( \frac{M}{6.2\times 10^{9}M_{\odot}}\right)\left(\frac{D}{16.9Mpc}\right)^{-1} \mu \text{as}.
\end{equation}

For the M87* black hole, the angular size of the photon ring, corresponding to the strong deflection limit, enables the placement of constraints on the bumblebee parameter $\ell$ using EHT data \cite{EventHorizonTelescope:2019dse}. The size of the photon ring for M87* is measured within the range $18.5 \ \mu\text{as} < \theta_{\text{p}} < 21 \ \mu\text{as}$. Similarly, for Sgr A*, the photon ring radius is found to be within $22 \ \mu\text{as} < \theta_{\text{p}} < 32.5 \ \mu\text{as}$, as reported in \cite{EventHorizonTelescope:2022wkp}. These measurements contribute significantly to establishing limits on the bumblebee parameter $\ell$. Upon application of the EHT data, it was shown consistently that the expected value of the bumblebee parameter is at the $-10^{26}$ order of magnitude. Considering the range given for the photon ring for both M87* and Sgr A*, the value of the bumblebee parameter is consistent. This is in line with the behavior given in Fig. \ref{deflectionangle2}. It shows that in the strong field region, the bumblebee parameter converges.

\textcolor{black}{Comparing this to the Schwarzschild case, our results retain the characteristic logarithmic divergence of the deflection angle near the photon sphere, as seen in Schwarzschild and Kerr spacetimes. However, the presence of the bumblebee parameter $\ell$ modifies the critical impact parameter and the deflection coefficients. Specifically, for positive $\ell$, the deflection angle is larger than in the Schwarzschild case, indicating stronger lensing effects, while for negative $\ell$, the deflection angle is reduced. Compared to Kerr black holes, the deviations induced by $\ell$ are quantitatively similar to those caused by spin, though arising from a different physical mechanism. These results reveal the potential of strong gravitational lensing as a tool for testing deviations from GR.}

\textcolor{black}{As a final remark, a comparison between the black hole solution presented in this study and the one in Ref. \cite{Filho:2022yrk} highlights key differences in the treatment of LSB and its observational consequences. The black hole model in Ref. \cite{Filho:2022yrk} is derived within a metric-affine extension of the Standard Model Extension and features a modified Schwarzschild-like metric where the effects of LSB appear through the coefficient $ X = \xi b^2 $. In contrast, the present study investigates a black hole in Einstein-Bumblebee gravity, which also incorporates LSB but introduces an effective cosmological constant $ \Lambda_e $, leading to distinct thermodynamic and observational properties. Both models predict deviations from GR, particularly in terms of modifications to light deflection, black hole shadows, and astrophysical observables. However, the presence of a cosmological constant in Einstein-Bumblebee gravity influences the black hole's shadow radius and deflection angle, offering additional constraints when compared to the metric-affine bumblebee model. While both solutions suggest measurable departures from GR, the role of $ \Lambda_e $ in the current work introduces an additional scale that affects the observational parameter space for testing LSB effects. Future high-precision astronomical observations, such as those from the EHT, may provide a means to distinguish between these two different formulations of Lorentz-violating gravity.}

\section{Conclusions} \label{conc}
In this study, we explored the effects of the cosmological constant and the bumblebee field on the thermodynamic and optical properties of a non-asymptotically flat black hole in Einstein-Bumblebee gravity. The horizon structure, depicted in Fig. \ref{structure Horizon}, reveals a single horizon for AdS spacetimes and dual horizons (black hole and cosmological) for dS spacetimes, with the bumblebee parameter and cosmological constant significantly influencing their radii, as shown in Figs. \ref{dSBHH}-\ref{Dplot} for dS and Fig. \ref{AdSBHH} for AdS cases. In the AdS context, the Hawking temperature increases as the cosmological constant decreases, with the bumblebee parameter modulating this effect, decreasing for $\ell < 0$ and increasing for $\ell > 0$, as illustrated in Fig. \ref{HawkingT}. The $P-v$ criticality analysis, visualized in Fig. \ref{PVT}, indicates that the critical temperature is independent of the bumblebee parameter, with a critical ratio of approximately $0.442$, exceeding the van der Waals value of $0.375$.

The black hole shadow analysis reveals modifications to the shadow radius driven by both the cosmological constant and the bumblebee parameter. Fig. \ref{shadow_cons} demonstrates constraints on the bumblebee parameter $\ell$ using EHT data for M87* and Sgr A*, showing a wide range for Sgr A* ($-10^{11} \lesssim \ell \lesssim 10^{11}$) and tighter bounds for M87* ($-34311 \lesssim \ell \lesssim 34309$). These deviations from GR predictions suggest that shadow observations can probe LSB. In the weak-field regime, Fig. \ref{deflectionangle} shows that the gravitational deflection angle increases with source distance for negative cosmological constants, with stronger effects for smaller $\Lambda_e$, and is enhanced for $\ell > 0$ compared to $\ell < 0$. In the strong-field regime, Fig. \ref{deflectionangle2} illustrates that the deflection angle increases with decreasing closest approach distance, with positive $\ell$ amplifying the effect and negative cosmological constants enhancing lensing compared to GR.

Our findings highlight the coupled influence of the cosmological constant ($\Lambda_e$) and bumblebee parameter ($\ell$) on black hole thermodynamics, shadow properties, and deflection angles. The interplay of these parameters, evident across Figs. \ref{structure Horizon}-\ref{deflectionangle2}, modifies observational signatures, offering potential probes for Lorentz symmetry violation through astrophysical observations. Given the cosmological constant's association with dark energy, its interaction with a Lorentz-violating background may have implications for cosmic expansion and structure formation, warranting further study.

Compared to prior work, our study extends Ref. \cite{Maluf:2021lwh} by incorporating EHT constraints and lensing analyses, revealing tighter bounds on $\ell$. Unlike the metric-affine Bumblebee model in Ref. \cite{Filho:2022yrk}, which uses $X = \xi b^2$ without $\Lambda_e$, our inclusion of the cosmological constant introduces additional observational signatures, particularly in shadow and strong lensing deviations. Recent studies, such as Refs. \cite{Gao:2024ejs,Lambiase:2024uzy}, focus on shadow and weak lensing in Bumblebee gravity, but our comprehensive analysis of both weak and strong deflection angles, coupled with $\Lambda_e$ effects, provides a more complete probe of Lorentz symmetry violation, aligning with EHT observations and offering a robust framework for future astrophysical tests.

Future research should extend this analysis to axisymmetric Kerr-Bumblebee-AdS spacetimes to capture spin-dependent effects on photon orbits and shadow deformation, critical for EHT comparisons. Full general-relativistic ray-tracing simulations, incorporating polarization and higher-order images, could refine constraints on $\ell$ and $\Lambda_e$ using high-resolution very-long-baseline interferometry data. Additionally, studying the linear perturbation spectrum and bumblebee field back-reaction on black hole evaporation will clarify the dynamical stability of Lorentz-violating black holes. Finally, embedding Einstein-Bumblebee gravity within cosmological perturbation theory could reveal novel imprints on large-scale structure and gravitational wave propagation, linking black hole phenomenology to tests of Lorentz symmetry and cosmic acceleration.

\section{Acknowledgements}
R. P. and A. \"O. would like to acknowledge networking support of the COST Action CA21106 - COSMIC WISPers in the Dark Universe: Theory, astrophysics and experiments (CosmicWISPers), the COST Action CA22113 - Fundamental challenges in theoretical physics (THEORY-CHALLENGES), the COST Action CA21136 - Addressing observational tensions in cosmology with systematics and fundamental physics (CosmoVerse), the COST Action CA23130 - Bridging high and low energies in search of quantum gravity (BridgeQG), and the COST Action CA23115 - Relativistic Quantum Information (RQI) funded by COST (European Cooperation in Science and Technology). A. \"O. also thanks to EMU, TUBITAK, ULAKBIM (Turkiye) and SCOAP3 (Switzerland) for their support.

\bibliography{references.bib}

\begin{thebibliography}{85}%
\makeatletter
\providecommand \@ifxundefined [1]{%
 \@ifx{#1\undefined}
}%
\providecommand \@ifnum [1]{%
 \ifnum #1\expandafter \@firstoftwo
 \else \expandafter \@secondoftwo
 \fi
}%
\providecommand \@ifx [1]{%
 \ifx #1\expandafter \@firstoftwo
 \else \expandafter \@secondoftwo
 \fi
}%
\providecommand \natexlab [1]{#1}%
\providecommand \enquote  [1]{``#1''}%
\providecommand \bibnamefont  [1]{#1}%
\providecommand \bibfnamefont [1]{#1}%
\providecommand \citenamefont [1]{#1}%
\providecommand \href@noop [0]{\@secondoftwo}%
\providecommand \href [0]{\begingroup \@sanitize@url \@href}%
\providecommand \@href[1]{\@@startlink{#1}\@@href}%
\providecommand \@@href[1]{\endgroup#1\@@endlink}%
\providecommand \@sanitize@url [0]{\catcode `\\12\catcode `\$12\catcode `\&12\catcode `\#12\catcode `\^12\catcode `\_12\catcode `\%12\relax}%
\providecommand \@@startlink[1]{}%
\providecommand \@@endlink[0]{}%
\providecommand \url  [0]{\begingroup\@sanitize@url \@url }%
\providecommand \@url [1]{\endgroup\@href {#1}{\urlprefix }}%
\providecommand \urlprefix  [0]{URL }%
\providecommand \Eprint [0]{\href }%
\providecommand \doibase [0]{http://dx.doi.org/}%
\providecommand \selectlanguage [0]{\@gobble}%
\providecommand \bibinfo  [0]{\@secondoftwo}%
\providecommand \bibfield  [0]{\@secondoftwo}%
\providecommand \translation [1]{[#1]}%
\providecommand \BibitemOpen [0]{}%
\providecommand \bibitemStop [0]{}%
\providecommand \bibitemNoStop [0]{.\EOS\space}%
\providecommand \EOS [0]{\spacefactor3000\relax}%
\providecommand \BibitemShut  [1]{\csname bibitem#1\endcsname}%
\let\auto@bib@innerbib\@empty
\bibitem [{\citenamefont {Einstein}(1915)}]{Einstein:1915ca}%
  \BibitemOpen
  \bibfield  {author} {\bibinfo {author} {\bibfnamefont {A.}~\bibnamefont {Einstein}},\ }\href@noop {} {\bibfield  {journal} {\bibinfo  {journal} {Sitzungsber. Preuss. Akad. Wiss. Berlin (Math. Phys. )}\ }\textbf {\bibinfo {volume} {1915}},\ \bibinfo {pages} {844} (\bibinfo {year} {1915})}\BibitemShut {NoStop}%
\bibitem [{\citenamefont {Wald}(1984)}]{Wald:1984rg}%
  \BibitemOpen
  \bibfield  {author} {\bibinfo {author} {\bibfnamefont {R.~M.}\ \bibnamefont {Wald}},\ }\href {\doibase 10.7208/chicago/9780226870373.001.0001} {\emph {\bibinfo {title} {{General Relativity}}}}\ (\bibinfo  {publisher} {Chicago Univ. Pr.},\ \bibinfo {address} {Chicago, USA},\ \bibinfo {year} {1984})\BibitemShut {NoStop}%
\bibitem [{\citenamefont {Chandrasekhar}(1998)}]{chandrasekhar1998mathematical}%
  \BibitemOpen
  \bibfield  {author} {\bibinfo {author} {\bibfnamefont {S.}~\bibnamefont {Chandrasekhar}},\ }\href@noop {} {\emph {\bibinfo {title} {The mathematical theory of black holes}}},\ Vol.~\bibinfo {volume} {69}\ (\bibinfo  {publisher} {Oxford university press},\ \bibinfo {year} {1998})\BibitemShut {NoStop}%
\bibitem [{\citenamefont {Abbott}\ \emph {et~al.}(2016)\citenamefont {Abbott} \emph {et~al.}}]{LIGOScientific:2016aoc}%
  \BibitemOpen
  \bibfield  {author} {\bibinfo {author} {\bibfnamefont {B.~P.}\ \bibnamefont {Abbott}} \emph {et~al.} (\bibinfo {collaboration} {LIGO Scientific, Virgo}),\ }\href {\doibase 10.1103/PhysRevLett.116.061102} {\bibfield  {journal} {\bibinfo  {journal} {Phys. Rev. Lett.}\ }\textbf {\bibinfo {volume} {116}},\ \bibinfo {pages} {061102} (\bibinfo {year} {2016})},\ \Eprint {http://arxiv.org/abs/1602.03837} {arXiv:1602.03837 [gr-qc]} \BibitemShut {NoStop}%
\bibitem [{\citenamefont {Abbott}\ \emph {et~al.}(2021)\citenamefont {Abbott} \emph {et~al.}}]{LIGOScientific:2021qlt}%
  \BibitemOpen
  \bibfield  {author} {\bibinfo {author} {\bibfnamefont {R.}~\bibnamefont {Abbott}} \emph {et~al.} (\bibinfo {collaboration} {LIGO Scientific, KAGRA, VIRGO}),\ }\href {\doibase 10.3847/2041-8213/ac082e} {\bibfield  {journal} {\bibinfo  {journal} {Astrophys. J. Lett.}\ }\textbf {\bibinfo {volume} {915}},\ \bibinfo {pages} {L5} (\bibinfo {year} {2021})},\ \Eprint {http://arxiv.org/abs/2106.15163} {arXiv:2106.15163 [astro-ph.HE]} \BibitemShut {NoStop}%
\bibitem [{\citenamefont {Akiyama}\ \emph {et~al.}(2022{\natexlab{a}})\citenamefont {Akiyama} \emph {et~al.}}]{EventHorizonTelescope:2022xqj}%
  \BibitemOpen
  \bibfield  {author} {\bibinfo {author} {\bibfnamefont {K.}~\bibnamefont {Akiyama}} \emph {et~al.} (\bibinfo {collaboration} {Event Horizon Telescope}),\ }\href {\doibase 10.3847/2041-8213/ac6756} {\bibfield  {journal} {\bibinfo  {journal} {Astrophys. J. Lett.}\ }\textbf {\bibinfo {volume} {930}},\ \bibinfo {pages} {L17} (\bibinfo {year} {2022}{\natexlab{a}})},\ \Eprint {http://arxiv.org/abs/2311.09484} {arXiv:2311.09484 [astro-ph.HE]} \BibitemShut {NoStop}%
\bibitem [{\citenamefont {Liberati}(2013)}]{Liberati:2013xla}%
  \BibitemOpen
  \bibfield  {author} {\bibinfo {author} {\bibfnamefont {S.}~\bibnamefont {Liberati}},\ }\href {\doibase 10.1088/0264-9381/30/13/133001} {\bibfield  {journal} {\bibinfo  {journal} {Class. Quant. Grav.}\ }\textbf {\bibinfo {volume} {30}},\ \bibinfo {pages} {133001} (\bibinfo {year} {2013})},\ \Eprint {http://arxiv.org/abs/1304.5795} {arXiv:1304.5795 [gr-qc]} \BibitemShut {NoStop}%
\bibitem [{\citenamefont {Kosteleck\'y}\ and\ \citenamefont {Lehnert}(2001)}]{PhysRevD.63.065008}%
  \BibitemOpen
  \bibfield  {author} {\bibinfo {author} {\bibfnamefont {V.~A.}\ \bibnamefont {Kosteleck\'y}}\ and\ \bibinfo {author} {\bibfnamefont {R.}~\bibnamefont {Lehnert}},\ }\href {\doibase 10.1103/PhysRevD.63.065008} {\bibfield  {journal} {\bibinfo  {journal} {Phys. Rev. D}\ }\textbf {\bibinfo {volume} {63}},\ \bibinfo {pages} {065008} (\bibinfo {year} {2001})}\BibitemShut {NoStop}%
\bibitem [{\citenamefont {Kosteleck\'y}\ and\ \citenamefont {Samuel}(1989)}]{PhysRevD.39.683}%
  \BibitemOpen
  \bibfield  {author} {\bibinfo {author} {\bibfnamefont {V.~A.}\ \bibnamefont {Kosteleck\'y}}\ and\ \bibinfo {author} {\bibfnamefont {S.}~\bibnamefont {Samuel}},\ }\href {\doibase 10.1103/PhysRevD.39.683} {\bibfield  {journal} {\bibinfo  {journal} {Phys. Rev. D}\ }\textbf {\bibinfo {volume} {39}},\ \bibinfo {pages} {683} (\bibinfo {year} {1989})}\BibitemShut {NoStop}%
\bibitem [{\citenamefont {Maluf}\ and\ \citenamefont {Neves}(2021{\natexlab{a}})}]{Maluf:2020kgf}%
  \BibitemOpen
  \bibfield  {author} {\bibinfo {author} {\bibfnamefont {R.~V.}\ \bibnamefont {Maluf}}\ and\ \bibinfo {author} {\bibfnamefont {J.~C.~S.}\ \bibnamefont {Neves}},\ }\href {\doibase 10.1103/PhysRevD.103.044002} {\bibfield  {journal} {\bibinfo  {journal} {Phys. Rev. D}\ }\textbf {\bibinfo {volume} {103}},\ \bibinfo {pages} {044002} (\bibinfo {year} {2021}{\natexlab{a}})},\ \Eprint {http://arxiv.org/abs/2011.12841} {arXiv:2011.12841 [gr-qc]} \BibitemShut {NoStop}%
\bibitem [{\citenamefont {Islam}\ \emph {et~al.}(2024)\citenamefont {Islam}, \citenamefont {Ghosh},\ and\ \citenamefont {Maharaj}}]{Islam:2024sph}%
  \BibitemOpen
  \bibfield  {author} {\bibinfo {author} {\bibfnamefont {S.~U.}\ \bibnamefont {Islam}}, \bibinfo {author} {\bibfnamefont {S.~G.}\ \bibnamefont {Ghosh}}, \ and\ \bibinfo {author} {\bibfnamefont {S.~D.}\ \bibnamefont {Maharaj}},\ }\href {\doibase 10.1088/1475-7516/2024/12/047} {\bibfield  {journal} {\bibinfo  {journal} {JCAP}\ }\textbf {\bibinfo {volume} {12}},\ \bibinfo {pages} {047} (\bibinfo {year} {2024})},\ \Eprint {http://arxiv.org/abs/2410.05395} {arXiv:2410.05395 [gr-qc]} \BibitemShut {NoStop}%
\bibitem [{\citenamefont {Afrin}\ \emph {et~al.}(2024)\citenamefont {Afrin}, \citenamefont {Ghosh},\ and\ \citenamefont {Wang}}]{Afrin:2024khy}%
  \BibitemOpen
  \bibfield  {author} {\bibinfo {author} {\bibfnamefont {M.}~\bibnamefont {Afrin}}, \bibinfo {author} {\bibfnamefont {S.~G.}\ \bibnamefont {Ghosh}}, \ and\ \bibinfo {author} {\bibfnamefont {A.}~\bibnamefont {Wang}},\ }\href {\doibase 10.1016/j.dark.2024.101642} {\bibfield  {journal} {\bibinfo  {journal} {Phys. Dark Univ.}\ }\textbf {\bibinfo {volume} {46}},\ \bibinfo {pages} {101642} (\bibinfo {year} {2024})},\ \Eprint {http://arxiv.org/abs/2409.06218} {arXiv:2409.06218 [gr-qc]} \BibitemShut {NoStop}%
\bibitem [{\citenamefont {Bekenstein}(2020)}]{bekenstein2020black}%
  \BibitemOpen
  \bibfield  {author} {\bibinfo {author} {\bibfnamefont {J.~D.}\ \bibnamefont {Bekenstein}},\ }in\ \href@noop {} {\emph {\bibinfo {booktitle} {JACOB BEKENSTEIN: The Conservative Revolutionary}}}\ (\bibinfo  {publisher} {World Scientific},\ \bibinfo {year} {2020})\ pp.\ \bibinfo {pages} {303--306}\BibitemShut {NoStop}%
\bibitem [{\citenamefont {Hawking}(1975)}]{hawking1975particle}%
  \BibitemOpen
  \bibfield  {author} {\bibinfo {author} {\bibfnamefont {S.~W.}\ \bibnamefont {Hawking}},\ }\href@noop {} {\bibfield  {journal} {\bibinfo  {journal} {Communications in mathematical physics}\ }\textbf {\bibinfo {volume} {43}},\ \bibinfo {pages} {199} (\bibinfo {year} {1975})}\BibitemShut {NoStop}%
\bibitem [{\citenamefont {Hawking}\ and\ \citenamefont {Page}(1983)}]{Hawking:1982dh}%
  \BibitemOpen
  \bibfield  {author} {\bibinfo {author} {\bibfnamefont {S.~W.}\ \bibnamefont {Hawking}}\ and\ \bibinfo {author} {\bibfnamefont {D.~N.}\ \bibnamefont {Page}},\ }\href {\doibase 10.1007/BF01208266} {\bibfield  {journal} {\bibinfo  {journal} {Commun. Math. Phys.}\ }\textbf {\bibinfo {volume} {87}},\ \bibinfo {pages} {577} (\bibinfo {year} {1983})}\BibitemShut {NoStop}%
\bibitem [{\citenamefont {Chamblin}\ \emph {et~al.}(1999)\citenamefont {Chamblin}, \citenamefont {Emparan}, \citenamefont {Johnson},\ and\ \citenamefont {Myers}}]{Chamblin:1999tk}%
  \BibitemOpen
  \bibfield  {author} {\bibinfo {author} {\bibfnamefont {A.}~\bibnamefont {Chamblin}}, \bibinfo {author} {\bibfnamefont {R.}~\bibnamefont {Emparan}}, \bibinfo {author} {\bibfnamefont {C.~V.}\ \bibnamefont {Johnson}}, \ and\ \bibinfo {author} {\bibfnamefont {R.~C.}\ \bibnamefont {Myers}},\ }\href {\doibase 10.1103/PhysRevD.60.064018} {\bibfield  {journal} {\bibinfo  {journal} {Phys. Rev. D}\ }\textbf {\bibinfo {volume} {60}},\ \bibinfo {pages} {064018} (\bibinfo {year} {1999})},\ \Eprint {http://arxiv.org/abs/hep-th/9902170} {arXiv:hep-th/9902170} \BibitemShut {NoStop}%
\bibitem [{\citenamefont {Kastor}\ \emph {et~al.}(2009)\citenamefont {Kastor}, \citenamefont {Ray},\ and\ \citenamefont {Traschen}}]{Kastor:2009wy}%
  \BibitemOpen
  \bibfield  {author} {\bibinfo {author} {\bibfnamefont {D.}~\bibnamefont {Kastor}}, \bibinfo {author} {\bibfnamefont {S.}~\bibnamefont {Ray}}, \ and\ \bibinfo {author} {\bibfnamefont {J.}~\bibnamefont {Traschen}},\ }\href {\doibase 10.1088/0264-9381/26/19/195011} {\bibfield  {journal} {\bibinfo  {journal} {Class. Quant. Grav.}\ }\textbf {\bibinfo {volume} {26}},\ \bibinfo {pages} {195011} (\bibinfo {year} {2009})},\ \Eprint {http://arxiv.org/abs/0904.2765} {arXiv:0904.2765 [hep-th]} \BibitemShut {NoStop}%
\bibitem [{\citenamefont {Banerjee}\ and\ \citenamefont {Roychowdhury}(2011)}]{Banerjee:2011au}%
  \BibitemOpen
  \bibfield  {author} {\bibinfo {author} {\bibfnamefont {R.}~\bibnamefont {Banerjee}}\ and\ \bibinfo {author} {\bibfnamefont {D.}~\bibnamefont {Roychowdhury}},\ }\href {\doibase 10.1007/JHEP11(2011)004} {\bibfield  {journal} {\bibinfo  {journal} {JHEP}\ }\textbf {\bibinfo {volume} {11}},\ \bibinfo {pages} {004} (\bibinfo {year} {2011})},\ \Eprint {http://arxiv.org/abs/1109.2433} {arXiv:1109.2433 [gr-qc]} \BibitemShut {NoStop}%
\bibitem [{\citenamefont {Gunasekaran}\ \emph {et~al.}(2012)\citenamefont {Gunasekaran}, \citenamefont {Mann},\ and\ \citenamefont {Kubiznak}}]{Gunasekaran:2012dq}%
  \BibitemOpen
  \bibfield  {author} {\bibinfo {author} {\bibfnamefont {S.}~\bibnamefont {Gunasekaran}}, \bibinfo {author} {\bibfnamefont {R.~B.}\ \bibnamefont {Mann}}, \ and\ \bibinfo {author} {\bibfnamefont {D.}~\bibnamefont {Kubiznak}},\ }\href {\doibase 10.1007/JHEP11(2012)110} {\bibfield  {journal} {\bibinfo  {journal} {JHEP}\ }\textbf {\bibinfo {volume} {11}},\ \bibinfo {pages} {110} (\bibinfo {year} {2012})},\ \Eprint {http://arxiv.org/abs/1208.6251} {arXiv:1208.6251 [hep-th]} \BibitemShut {NoStop}%
\bibitem [{\citenamefont {Zou}\ \emph {et~al.}(2014)\citenamefont {Zou}, \citenamefont {Zhang},\ and\ \citenamefont {Wang}}]{Zou:2013owa}%
  \BibitemOpen
  \bibfield  {author} {\bibinfo {author} {\bibfnamefont {D.-C.}\ \bibnamefont {Zou}}, \bibinfo {author} {\bibfnamefont {S.-J.}\ \bibnamefont {Zhang}}, \ and\ \bibinfo {author} {\bibfnamefont {B.}~\bibnamefont {Wang}},\ }\href {\doibase 10.1103/PhysRevD.89.044002} {\bibfield  {journal} {\bibinfo  {journal} {Phys. Rev. D}\ }\textbf {\bibinfo {volume} {89}},\ \bibinfo {pages} {044002} (\bibinfo {year} {2014})},\ \Eprint {http://arxiv.org/abs/1311.7299} {arXiv:1311.7299 [hep-th]} \BibitemShut {NoStop}%
\bibitem [{\citenamefont {Dolan}(2014)}]{Dolan:2014lea}%
  \BibitemOpen
  \bibfield  {author} {\bibinfo {author} {\bibfnamefont {B.~P.}\ \bibnamefont {Dolan}},\ }\href {\doibase 10.1088/0264-9381/31/16/165011} {\bibfield  {journal} {\bibinfo  {journal} {Class. Quant. Grav.}\ }\textbf {\bibinfo {volume} {31}},\ \bibinfo {pages} {165011} (\bibinfo {year} {2014})},\ \Eprint {http://arxiv.org/abs/1403.1507} {arXiv:1403.1507 [gr-qc]} \BibitemShut {NoStop}%
\bibitem [{\citenamefont {Promsiri}\ \emph {et~al.}(2020)\citenamefont {Promsiri}, \citenamefont {Hirunsirisawat},\ and\ \citenamefont {Liewrian}}]{Promsiri:2020jga}%
  \BibitemOpen
  \bibfield  {author} {\bibinfo {author} {\bibfnamefont {C.}~\bibnamefont {Promsiri}}, \bibinfo {author} {\bibfnamefont {E.}~\bibnamefont {Hirunsirisawat}}, \ and\ \bibinfo {author} {\bibfnamefont {W.}~\bibnamefont {Liewrian}},\ }\href {\doibase 10.1103/PhysRevD.102.064014} {\bibfield  {journal} {\bibinfo  {journal} {Phys. Rev. D}\ }\textbf {\bibinfo {volume} {102}},\ \bibinfo {pages} {064014} (\bibinfo {year} {2020})},\ \Eprint {http://arxiv.org/abs/2003.12986} {arXiv:2003.12986 [hep-th]} \BibitemShut {NoStop}%
\bibitem [{\citenamefont {Walia}\ \emph {et~al.}(2022)\citenamefont {Walia}, \citenamefont {Maharaj},\ and\ \citenamefont {Ghosh}}]{Walia:2021emv}%
  \BibitemOpen
  \bibfield  {author} {\bibinfo {author} {\bibfnamefont {R.~K.}\ \bibnamefont {Walia}}, \bibinfo {author} {\bibfnamefont {S.~D.}\ \bibnamefont {Maharaj}}, \ and\ \bibinfo {author} {\bibfnamefont {S.~G.}\ \bibnamefont {Ghosh}},\ }\href {\doibase 10.1140/epjc/s10052-022-10451-5} {\bibfield  {journal} {\bibinfo  {journal} {Eur. Phys. J. C}\ }\textbf {\bibinfo {volume} {82}},\ \bibinfo {pages} {547} (\bibinfo {year} {2022})},\ \Eprint {http://arxiv.org/abs/2109.08055} {arXiv:2109.08055 [gr-qc]} \BibitemShut {NoStop}%
\bibitem [{\citenamefont {Ali}\ \emph {et~al.}(2023)\citenamefont {Ali}, \citenamefont {Ghosh},\ and\ \citenamefont {Wang}}]{Ali:2023ppg}%
  \BibitemOpen
  \bibfield  {author} {\bibinfo {author} {\bibfnamefont {M.~S.}\ \bibnamefont {Ali}}, \bibinfo {author} {\bibfnamefont {S.~G.}\ \bibnamefont {Ghosh}}, \ and\ \bibinfo {author} {\bibfnamefont {A.}~\bibnamefont {Wang}},\ }\href {\doibase 10.1103/PhysRevD.108.044045} {\bibfield  {journal} {\bibinfo  {journal} {Phys. Rev. D}\ }\textbf {\bibinfo {volume} {108}},\ \bibinfo {pages} {044045} (\bibinfo {year} {2023})},\ \Eprint {http://arxiv.org/abs/2308.00489} {arXiv:2308.00489 [gr-qc]} \BibitemShut {NoStop}%
\bibitem [{\citenamefont {Cano}\ and\ \citenamefont {David}(2024)}]{Cano:2024tcr}%
  \BibitemOpen
  \bibfield  {author} {\bibinfo {author} {\bibfnamefont {P.~A.}\ \bibnamefont {Cano}}\ and\ \bibinfo {author} {\bibfnamefont {M.}~\bibnamefont {David}},\ }\href {\doibase 10.1007/JHEP03(2024)036} {\bibfield  {journal} {\bibinfo  {journal} {JHEP}\ }\textbf {\bibinfo {volume} {03}},\ \bibinfo {pages} {036} (\bibinfo {year} {2024})},\ \Eprint {http://arxiv.org/abs/2402.02215} {arXiv:2402.02215 [hep-th]} \BibitemShut {NoStop}%
\bibitem [{\citenamefont {Kubiz{\v{n}}{\'a}k}\ and\ \citenamefont {Mann}(2012)}]{kubizvnak2012p}%
  \BibitemOpen
  \bibfield  {author} {\bibinfo {author} {\bibfnamefont {D.}~\bibnamefont {Kubiz{\v{n}}{\'a}k}}\ and\ \bibinfo {author} {\bibfnamefont {R.~B.}\ \bibnamefont {Mann}},\ }\href@noop {} {\bibfield  {journal} {\bibinfo  {journal} {Journal of High Energy Physics}\ }\textbf {\bibinfo {volume} {2012}},\ \bibinfo {pages} {1} (\bibinfo {year} {2012})}\BibitemShut {NoStop}%
\bibitem [{\citenamefont {Haditale}\ and\ \citenamefont {Malekolkalami}(2024)}]{Haditale:2023adr}%
  \BibitemOpen
  \bibfield  {author} {\bibinfo {author} {\bibfnamefont {M.}~\bibnamefont {Haditale}}\ and\ \bibinfo {author} {\bibfnamefont {B.}~\bibnamefont {Malekolkalami}},\ }\href {\doibase 10.1002/prop.202300267} {\bibfield  {journal} {\bibinfo  {journal} {Fortsch. Phys.}\ }\textbf {\bibinfo {volume} {72}},\ \bibinfo {pages} {2300267} (\bibinfo {year} {2024})},\ \Eprint {http://arxiv.org/abs/2308.16627} {arXiv:2308.16627 [gr-qc]} \BibitemShut {NoStop}%
\bibitem [{\citenamefont {Jafarzade}\ \emph {et~al.}(2024)\citenamefont {Jafarzade}, \citenamefont {Panah},\ and\ \citenamefont {Rodrigues}}]{jafarzade2024thermodynamics}%
  \BibitemOpen
  \bibfield  {author} {\bibinfo {author} {\bibfnamefont {K.}~\bibnamefont {Jafarzade}}, \bibinfo {author} {\bibfnamefont {B.~E.}\ \bibnamefont {Panah}}, \ and\ \bibinfo {author} {\bibfnamefont {M.}~\bibnamefont {Rodrigues}},\ }\href@noop {} {\bibfield  {journal} {\bibinfo  {journal} {Classical and Quantum Gravity}\ }\textbf {\bibinfo {volume} {41}},\ \bibinfo {pages} {065007} (\bibinfo {year} {2024})}\BibitemShut {NoStop}%
\bibitem [{\citenamefont {Wang}\ \emph {et~al.}(2024)\citenamefont {Wang}, \citenamefont {Ma}, \citenamefont {You}, \citenamefont {Tang}, \citenamefont {Feng}, \citenamefont {Hu},\ and\ \citenamefont {Deng}}]{wang2024thermodynamics}%
  \BibitemOpen
  \bibfield  {author} {\bibinfo {author} {\bibfnamefont {R.-B.}\ \bibnamefont {Wang}}, \bibinfo {author} {\bibfnamefont {S.-J.}\ \bibnamefont {Ma}}, \bibinfo {author} {\bibfnamefont {L.}~\bibnamefont {You}}, \bibinfo {author} {\bibfnamefont {Y.-C.}\ \bibnamefont {Tang}}, \bibinfo {author} {\bibfnamefont {Y.-H.}\ \bibnamefont {Feng}}, \bibinfo {author} {\bibfnamefont {X.-R.}\ \bibnamefont {Hu}}, \ and\ \bibinfo {author} {\bibfnamefont {J.-B.}\ \bibnamefont {Deng}},\ }\href {\doibase 10.1140/epjc/s10052-024-13505-y} {\bibfield  {journal} {\bibinfo  {journal} {The European Physical Journal C}\ }\textbf {\bibinfo {volume} {84}},\ \bibinfo {pages} {1161} (\bibinfo {year} {2024})}\BibitemShut {NoStop}%
\bibitem [{\citenamefont {Synge}(1966)}]{Synge:1966okc}%
  \BibitemOpen
  \bibfield  {author} {\bibinfo {author} {\bibfnamefont {J.~L.}\ \bibnamefont {Synge}},\ }\href {\doibase 10.1093/mnras/131.3.463} {\bibfield  {journal} {\bibinfo  {journal} {Mon. Not. Roy. Astron. Soc.}\ }\textbf {\bibinfo {volume} {131}},\ \bibinfo {pages} {463} (\bibinfo {year} {1966})}\BibitemShut {NoStop}%
\bibitem [{\citenamefont {Zakharov}(2024)}]{Zakharov:2023yjl}%
  \BibitemOpen
  \bibfield  {author} {\bibinfo {author} {\bibfnamefont {A.~F.}\ \bibnamefont {Zakharov}},\ }\href {\doibase 10.1142/S0218271823400047} {\bibfield  {journal} {\bibinfo  {journal} {Int. J. Mod. Phys. D}\ }\textbf {\bibinfo {volume} {33}},\ \bibinfo {pages} {2340004} (\bibinfo {year} {2024})},\ \Eprint {http://arxiv.org/abs/2308.01301} {arXiv:2308.01301 [gr-qc]} \BibitemShut {NoStop}%
\bibitem [{\citenamefont {Luminet}(1979)}]{Luminet:1979nyg}%
  \BibitemOpen
  \bibfield  {author} {\bibinfo {author} {\bibfnamefont {J.~P.}\ \bibnamefont {Luminet}},\ }\href@noop {} {\bibfield  {journal} {\bibinfo  {journal} {Astron. Astrophys.}\ }\textbf {\bibinfo {volume} {75}},\ \bibinfo {pages} {228} (\bibinfo {year} {1979})}\BibitemShut {NoStop}%
\bibitem [{\citenamefont {Falcke}\ \emph {et~al.}(2000)\citenamefont {Falcke}, \citenamefont {Melia},\ and\ \citenamefont {Agol}}]{Falcke:1999pj}%
  \BibitemOpen
  \bibfield  {author} {\bibinfo {author} {\bibfnamefont {H.}~\bibnamefont {Falcke}}, \bibinfo {author} {\bibfnamefont {F.}~\bibnamefont {Melia}}, \ and\ \bibinfo {author} {\bibfnamefont {E.}~\bibnamefont {Agol}},\ }\href {\doibase 10.1086/312423} {\bibfield  {journal} {\bibinfo  {journal} {Astrophys. J. Lett.}\ }\textbf {\bibinfo {volume} {528}},\ \bibinfo {pages} {L13} (\bibinfo {year} {2000})},\ \Eprint {http://arxiv.org/abs/astro-ph/9912263} {arXiv:astro-ph/9912263} \BibitemShut {NoStop}%
\bibitem [{\citenamefont {Claudel}\ \emph {et~al.}(2001)\citenamefont {Claudel}, \citenamefont {Virbhadra},\ and\ \citenamefont {Ellis}}]{Claudel:2000yi}%
  \BibitemOpen
  \bibfield  {author} {\bibinfo {author} {\bibfnamefont {C.-M.}\ \bibnamefont {Claudel}}, \bibinfo {author} {\bibfnamefont {K.~S.}\ \bibnamefont {Virbhadra}}, \ and\ \bibinfo {author} {\bibfnamefont {G.~F.~R.}\ \bibnamefont {Ellis}},\ }\href {\doibase 10.1063/1.1308507} {\bibfield  {journal} {\bibinfo  {journal} {J. Math. Phys.}\ }\textbf {\bibinfo {volume} {42}},\ \bibinfo {pages} {818} (\bibinfo {year} {2001})},\ \Eprint {http://arxiv.org/abs/gr-qc/0005050} {arXiv:gr-qc/0005050} \BibitemShut {NoStop}%
\bibitem [{\citenamefont {{Cunningham}}\ and\ \citenamefont {{Bardeen}}(1973)}]{Cunningham}%
  \BibitemOpen
  \bibfield  {author} {\bibinfo {author} {\bibfnamefont {C.~T.}\ \bibnamefont {{Cunningham}}}\ and\ \bibinfo {author} {\bibfnamefont {J.~M.}\ \bibnamefont {{Bardeen}}},\ }\href {\doibase 10.1086/152223} {\bibfield  {journal} {\bibinfo  {journal} {\apj}\ }\textbf {\bibinfo {volume} {183}},\ \bibinfo {pages} {237} (\bibinfo {year} {1973})}\BibitemShut {NoStop}%
\bibitem [{\citenamefont {{Bardeen}}(1974)}]{1974IAUS...64..132B}%
  \BibitemOpen
  \bibfield  {author} {\bibinfo {author} {\bibfnamefont {J.~M.}\ \bibnamefont {{Bardeen}}},\ }in\ \href@noop {} {\emph {\bibinfo {booktitle} {Gravitational Radiation and Gravitational Collapse}}},\ Vol.~\bibinfo {volume} {64},\ \bibinfo {editor} {edited by\ \bibinfo {editor} {\bibfnamefont {C.}~\bibnamefont {{Dewitt-Morette}}}}\ (\bibinfo  {publisher} {Springer Dordrecht},\ \bibinfo {year} {1974})\ p.\ \bibinfo {pages} {132}\BibitemShut {NoStop}%
\bibitem [{\citenamefont {Akiyama}\ \emph {et~al.}(2019{\natexlab{a}})\citenamefont {Akiyama} \emph {et~al.}}]{EventHorizonTelescope:2019dse}%
  \BibitemOpen
  \bibfield  {author} {\bibinfo {author} {\bibfnamefont {K.}~\bibnamefont {Akiyama}} \emph {et~al.} (\bibinfo {collaboration} {Event Horizon Telescope}),\ }\href {\doibase 10.3847/2041-8213/ab0ec7} {\bibfield  {journal} {\bibinfo  {journal} {Astrophys. J. Lett.}\ }\textbf {\bibinfo {volume} {875}},\ \bibinfo {pages} {L1} (\bibinfo {year} {2019}{\natexlab{a}})},\ \Eprint {http://arxiv.org/abs/1906.11238} {arXiv:1906.11238 [astro-ph.GA]} \BibitemShut {NoStop}%
\bibitem [{\citenamefont {Akiyama}\ \emph {et~al.}(2019{\natexlab{b}})\citenamefont {Akiyama} \emph {et~al.}}]{EventHorizonTelescope:2019ths}%
  \BibitemOpen
  \bibfield  {author} {\bibinfo {author} {\bibfnamefont {K.}~\bibnamefont {Akiyama}} \emph {et~al.} (\bibinfo {collaboration} {Event Horizon Telescope}),\ }\href {\doibase 10.3847/2041-8213/ab0e85} {\bibfield  {journal} {\bibinfo  {journal} {Astrophys. J. Lett.}\ }\textbf {\bibinfo {volume} {875}},\ \bibinfo {pages} {L4} (\bibinfo {year} {2019}{\natexlab{b}})},\ \Eprint {http://arxiv.org/abs/1906.11241} {arXiv:1906.11241 [astro-ph.GA]} \BibitemShut {NoStop}%
\bibitem [{\citenamefont {Zakharov}\ \emph {et~al.}(2012)\citenamefont {Zakharov}, \citenamefont {De~Paolis}, \citenamefont {Ingrosso},\ and\ \citenamefont {Nucita}}]{Zakharov:2011zz}%
  \BibitemOpen
  \bibfield  {author} {\bibinfo {author} {\bibfnamefont {A.~F.}\ \bibnamefont {Zakharov}}, \bibinfo {author} {\bibfnamefont {F.}~\bibnamefont {De~Paolis}}, \bibinfo {author} {\bibfnamefont {G.}~\bibnamefont {Ingrosso}}, \ and\ \bibinfo {author} {\bibfnamefont {A.~A.}\ \bibnamefont {Nucita}},\ }\href {\doibase 10.1016/j.newar.2011.09.002} {\bibfield  {journal} {\bibinfo  {journal} {New Astron. Rev.}\ }\textbf {\bibinfo {volume} {56}},\ \bibinfo {pages} {64} (\bibinfo {year} {2012})}\BibitemShut {NoStop}%
\bibitem [{\citenamefont {Zakharov}\ \emph {et~al.}(2005)\citenamefont {Zakharov}, \citenamefont {De~Paolis}, \citenamefont {Ingrosso},\ and\ \citenamefont {Nucita}}]{Zakharov:2005ek}%
  \BibitemOpen
  \bibfield  {author} {\bibinfo {author} {\bibfnamefont {A.~F.}\ \bibnamefont {Zakharov}}, \bibinfo {author} {\bibfnamefont {F.}~\bibnamefont {De~Paolis}}, \bibinfo {author} {\bibfnamefont {G.}~\bibnamefont {Ingrosso}}, \ and\ \bibinfo {author} {\bibfnamefont {A.~A.}\ \bibnamefont {Nucita}},\ }\href {\doibase 10.1051/0004-6361:20053432} {\bibfield  {journal} {\bibinfo  {journal} {Astron. Astrophys.}\ }\textbf {\bibinfo {volume} {442}},\ \bibinfo {pages} {795} (\bibinfo {year} {2005})},\ \Eprint {http://arxiv.org/abs/astro-ph/0505286} {arXiv:astro-ph/0505286} \BibitemShut {NoStop}%
\bibitem [{\citenamefont {Kocherlakota}\ \emph {et~al.}(2021)\citenamefont {Kocherlakota} \emph {et~al.}}]{EventHorizonTelescope:2021dqv}%
  \BibitemOpen
  \bibfield  {author} {\bibinfo {author} {\bibfnamefont {P.}~\bibnamefont {Kocherlakota}} \emph {et~al.} (\bibinfo {collaboration} {Event Horizon Telescope}),\ }\href {\doibase 10.1103/PhysRevD.103.104047} {\bibfield  {journal} {\bibinfo  {journal} {Phys. Rev. D}\ }\textbf {\bibinfo {volume} {103}},\ \bibinfo {pages} {104047} (\bibinfo {year} {2021})},\ \Eprint {http://arxiv.org/abs/2105.09343} {arXiv:2105.09343 [gr-qc]} \BibitemShut {NoStop}%
\bibitem [{\citenamefont {Gao}(2024)}]{Gao:2024ejs}%
  \BibitemOpen
  \bibfield  {author} {\bibinfo {author} {\bibfnamefont {X.-J.}\ \bibnamefont {Gao}},\ }\href {\doibase 10.1140/epjc/s10052-024-13338-9} {\bibfield  {journal} {\bibinfo  {journal} {Eur. Phys. J. C}\ }\textbf {\bibinfo {volume} {84}},\ \bibinfo {pages} {973} (\bibinfo {year} {2024})},\ \Eprint {http://arxiv.org/abs/2409.12531} {arXiv:2409.12531 [gr-qc]} \BibitemShut {NoStop}%
\bibitem [{\citenamefont {Lambiase}\ \emph {et~al.}(2024)\citenamefont {Lambiase}, \citenamefont {Pantig},\ and\ \citenamefont {\"Ovg\"un}}]{Lambiase:2024uzy}%
  \BibitemOpen
  \bibfield  {author} {\bibinfo {author} {\bibfnamefont {G.}~\bibnamefont {Lambiase}}, \bibinfo {author} {\bibfnamefont {R.~C.}\ \bibnamefont {Pantig}}, \ and\ \bibinfo {author} {\bibfnamefont {A.}~\bibnamefont {\"Ovg\"un}},\ }\href {\doibase 10.1209/0295-5075/ad8d79} {\bibfield  {journal} {\bibinfo  {journal} {EPL}\ }\textbf {\bibinfo {volume} {148}},\ \bibinfo {pages} {49001} (\bibinfo {year} {2024})},\ \Eprint {http://arxiv.org/abs/2408.09620} {arXiv:2408.09620 [gr-qc]} \BibitemShut {NoStop}%
\bibitem [{\citenamefont {Li}\ \emph {et~al.}(2024)\citenamefont {Li}, \citenamefont {Zhang},\ and\ \citenamefont {Huang}}]{Li:2024owp}%
  \BibitemOpen
  \bibfield  {author} {\bibinfo {author} {\bibfnamefont {H.-L.}\ \bibnamefont {Li}}, \bibinfo {author} {\bibfnamefont {M.}~\bibnamefont {Zhang}}, \ and\ \bibinfo {author} {\bibfnamefont {Y.-M.}\ \bibnamefont {Huang}},\ }\href {\doibase 10.1140/epjc/s10052-024-13194-7} {\bibfield  {journal} {\bibinfo  {journal} {Eur. Phys. J. C}\ }\textbf {\bibinfo {volume} {84}},\ \bibinfo {pages} {860} (\bibinfo {year} {2024})}\BibitemShut {NoStop}%
\bibitem [{\citenamefont {Atamurotov}\ \emph {et~al.}(2024)\citenamefont {Atamurotov}, \citenamefont {Sarikulov}, \citenamefont {Ghosh},\ and\ \citenamefont {Mustafa}}]{Atamurotov:2024nre}%
  \BibitemOpen
  \bibfield  {author} {\bibinfo {author} {\bibfnamefont {F.}~\bibnamefont {Atamurotov}}, \bibinfo {author} {\bibfnamefont {F.}~\bibnamefont {Sarikulov}}, \bibinfo {author} {\bibfnamefont {S.~G.}\ \bibnamefont {Ghosh}}, \ and\ \bibinfo {author} {\bibfnamefont {G.}~\bibnamefont {Mustafa}},\ }\href {\doibase 10.1016/j.dark.2024.101625} {\bibfield  {journal} {\bibinfo  {journal} {Phys. Dark Univ.}\ }\textbf {\bibinfo {volume} {46}},\ \bibinfo {pages} {101625} (\bibinfo {year} {2024})}\BibitemShut {NoStop}%
\bibitem [{\citenamefont {Chen}\ \emph {et~al.}(2024)\citenamefont {Chen}, \citenamefont {Dong}, \citenamefont {Maghsoodi}, \citenamefont {Hassanabadi}, \citenamefont {K\v{r}i\v{z}}, \citenamefont {Zare},\ and\ \citenamefont {Hassanabadi}}]{Chen:2024mlr}%
  \BibitemOpen
  \bibfield  {author} {\bibinfo {author} {\bibfnamefont {H.}~\bibnamefont {Chen}}, \bibinfo {author} {\bibfnamefont {S.~H.}\ \bibnamefont {Dong}}, \bibinfo {author} {\bibfnamefont {E.}~\bibnamefont {Maghsoodi}}, \bibinfo {author} {\bibfnamefont {S.}~\bibnamefont {Hassanabadi}}, \bibinfo {author} {\bibfnamefont {J.}~\bibnamefont {K\v{r}i\v{z}}}, \bibinfo {author} {\bibfnamefont {S.}~\bibnamefont {Zare}}, \ and\ \bibinfo {author} {\bibfnamefont {H.}~\bibnamefont {Hassanabadi}},\ }\href {\doibase 10.1140/epjp/s13360-024-05561-w} {\bibfield  {journal} {\bibinfo  {journal} {Eur. Phys. J. Plus}\ }\textbf {\bibinfo {volume} {139}},\ \bibinfo {pages} {759} (\bibinfo {year} {2024})}\BibitemShut {NoStop}%
\bibitem [{\citenamefont {Papnoi}\ \emph {et~al.}(2024)\citenamefont {Papnoi}, \citenamefont {Atamurotov}, \citenamefont {Nandan}, \citenamefont {Pandey}, \citenamefont {Mustafa},\ and\ \citenamefont {Saidov}}]{Papnoi:2024wzs}%
  \BibitemOpen
  \bibfield  {author} {\bibinfo {author} {\bibfnamefont {U.}~\bibnamefont {Papnoi}}, \bibinfo {author} {\bibfnamefont {F.}~\bibnamefont {Atamurotov}}, \bibinfo {author} {\bibfnamefont {H.}~\bibnamefont {Nandan}}, \bibinfo {author} {\bibfnamefont {P.}~\bibnamefont {Pandey}}, \bibinfo {author} {\bibfnamefont {G.}~\bibnamefont {Mustafa}}, \ and\ \bibinfo {author} {\bibfnamefont {I.}~\bibnamefont {Saidov}},\ }\href {\doibase 10.1016/j.dark.2024.101612} {\bibfield  {journal} {\bibinfo  {journal} {Phys. Dark Univ.}\ }\textbf {\bibinfo {volume} {46}},\ \bibinfo {pages} {101612} (\bibinfo {year} {2024})}\BibitemShut {NoStop}%
\bibitem [{\citenamefont {Yildiz}\ \emph {et~al.}(2024)\citenamefont {Yildiz}, \citenamefont {Ditta}, \citenamefont {Ashraf}, \citenamefont {G\"udekli}, \citenamefont {Alanazi},\ and\ \citenamefont {Reyimberganov}}]{Yildiz:2024dkt}%
  \BibitemOpen
  \bibfield  {author} {\bibinfo {author} {\bibfnamefont {G.~D.~A.}\ \bibnamefont {Yildiz}}, \bibinfo {author} {\bibfnamefont {A.}~\bibnamefont {Ditta}}, \bibinfo {author} {\bibfnamefont {A.}~\bibnamefont {Ashraf}}, \bibinfo {author} {\bibfnamefont {E.}~\bibnamefont {G\"udekli}}, \bibinfo {author} {\bibfnamefont {Y.~M.}\ \bibnamefont {Alanazi}}, \ and\ \bibinfo {author} {\bibfnamefont {A.}~\bibnamefont {Reyimberganov}},\ }\href {\doibase 10.1016/j.dark.2024.101583} {\bibfield  {journal} {\bibinfo  {journal} {Phys. Dark Univ.}\ }\textbf {\bibinfo {volume} {46}},\ \bibinfo {pages} {101583} (\bibinfo {year} {2024})}\BibitemShut {NoStop}%
\bibitem [{\citenamefont {Zare}\ \emph {et~al.}(2024)\citenamefont {Zare}, \citenamefont {Nieto}, \citenamefont {Feng}, \citenamefont {Dong},\ and\ \citenamefont {Hassanabadi}}]{Zare:2024dtf}%
  \BibitemOpen
  \bibfield  {author} {\bibinfo {author} {\bibfnamefont {S.}~\bibnamefont {Zare}}, \bibinfo {author} {\bibfnamefont {L.~M.}\ \bibnamefont {Nieto}}, \bibinfo {author} {\bibfnamefont {X.-H.}\ \bibnamefont {Feng}}, \bibinfo {author} {\bibfnamefont {S.-H.}\ \bibnamefont {Dong}}, \ and\ \bibinfo {author} {\bibfnamefont {H.}~\bibnamefont {Hassanabadi}},\ }\href {\doibase 10.1088/1475-7516/2024/08/041} {\bibfield  {journal} {\bibinfo  {journal} {JCAP}\ }\textbf {\bibinfo {volume} {08}},\ \bibinfo {pages} {041} (\bibinfo {year} {2024})},\ \Eprint {http://arxiv.org/abs/2406.07300} {arXiv:2406.07300 [astro-ph.HE]} \BibitemShut {NoStop}%
\bibitem [{\citenamefont {Refsdal}(1964)}]{Refsdal:1964yk}%
  \BibitemOpen
  \bibfield  {author} {\bibinfo {author} {\bibfnamefont {S.}~\bibnamefont {Refsdal}},\ }\href@noop {} {\bibfield  {journal} {\bibinfo  {journal} {Mon. Not. Roy. Astron. Soc.}\ }\textbf {\bibinfo {volume} {128}},\ \bibinfo {pages} {295} (\bibinfo {year} {1964})}\BibitemShut {NoStop}%
\bibitem [{\citenamefont {Virbhadra}\ and\ \citenamefont {Ellis}(2000)}]{PhysRevD.62.084003}%
  \BibitemOpen
  \bibfield  {author} {\bibinfo {author} {\bibfnamefont {K.~S.}\ \bibnamefont {Virbhadra}}\ and\ \bibinfo {author} {\bibfnamefont {G.~F.~R.}\ \bibnamefont {Ellis}},\ }\href {\doibase 10.1103/PhysRevD.62.084003} {\bibfield  {journal} {\bibinfo  {journal} {Phys. Rev. D}\ }\textbf {\bibinfo {volume} {62}},\ \bibinfo {pages} {084003} (\bibinfo {year} {2000})}\BibitemShut {NoStop}%
\bibitem [{\citenamefont {Gibbons}\ and\ \citenamefont {Werner}(2008)}]{Gibbons_2008}%
  \BibitemOpen
  \bibfield  {author} {\bibinfo {author} {\bibfnamefont {G.~W.}\ \bibnamefont {Gibbons}}\ and\ \bibinfo {author} {\bibfnamefont {M.~C.}\ \bibnamefont {Werner}},\ }\href {\doibase 10.1088/0264-9381/25/23/235009} {\bibfield  {journal} {\bibinfo  {journal} {Classical and Quantum Gravity}\ }\textbf {\bibinfo {volume} {25}},\ \bibinfo {pages} {235009} (\bibinfo {year} {2008})}\BibitemShut {NoStop}%
\bibitem [{\citenamefont {Werner}(2012)}]{Werner:2012rc}%
  \BibitemOpen
  \bibfield  {author} {\bibinfo {author} {\bibfnamefont {M.~C.}\ \bibnamefont {Werner}},\ }\href {\doibase 10.1007/s10714-012-1458-9} {\bibfield  {journal} {\bibinfo  {journal} {Gen. Rel. Grav.}\ }\textbf {\bibinfo {volume} {44}},\ \bibinfo {pages} {3047} (\bibinfo {year} {2012})},\ \Eprint {http://arxiv.org/abs/1205.3876} {arXiv:1205.3876 [gr-qc]} \BibitemShut {NoStop}%
\bibitem [{\citenamefont {Ishihara}\ \emph {et~al.}(2016)\citenamefont {Ishihara}, \citenamefont {Suzuki}, \citenamefont {Ono}, \citenamefont {Kitamura},\ and\ \citenamefont {Asada}}]{PhysRevD.94.084015}%
  \BibitemOpen
  \bibfield  {author} {\bibinfo {author} {\bibfnamefont {A.}~\bibnamefont {Ishihara}}, \bibinfo {author} {\bibfnamefont {Y.}~\bibnamefont {Suzuki}}, \bibinfo {author} {\bibfnamefont {T.}~\bibnamefont {Ono}}, \bibinfo {author} {\bibfnamefont {T.}~\bibnamefont {Kitamura}}, \ and\ \bibinfo {author} {\bibfnamefont {H.}~\bibnamefont {Asada}},\ }\href {\doibase 10.1103/PhysRevD.94.084015} {\bibfield  {journal} {\bibinfo  {journal} {Phys. Rev. D}\ }\textbf {\bibinfo {volume} {94}},\ \bibinfo {pages} {084015} (\bibinfo {year} {2016})}\BibitemShut {NoStop}%
\bibitem [{\citenamefont {Ishihara}\ \emph {et~al.}(2017)\citenamefont {Ishihara}, \citenamefont {Suzuki}, \citenamefont {Ono},\ and\ \citenamefont {Asada}}]{PhysRevD.95.044017}%
  \BibitemOpen
  \bibfield  {author} {\bibinfo {author} {\bibfnamefont {A.}~\bibnamefont {Ishihara}}, \bibinfo {author} {\bibfnamefont {Y.}~\bibnamefont {Suzuki}}, \bibinfo {author} {\bibfnamefont {T.}~\bibnamefont {Ono}}, \ and\ \bibinfo {author} {\bibfnamefont {H.}~\bibnamefont {Asada}},\ }\href {\doibase 10.1103/PhysRevD.95.044017} {\bibfield  {journal} {\bibinfo  {journal} {Phys. Rev. D}\ }\textbf {\bibinfo {volume} {95}},\ \bibinfo {pages} {044017} (\bibinfo {year} {2017})}\BibitemShut {NoStop}%
\bibitem [{\citenamefont {Jusufi}\ \emph {et~al.}(2017)\citenamefont {Jusufi}, \citenamefont {Werner}, \citenamefont {Banerjee},\ and\ \citenamefont {\"Ovg\"un}}]{PhysRevD.95.104012}%
  \BibitemOpen
  \bibfield  {author} {\bibinfo {author} {\bibfnamefont {K.}~\bibnamefont {Jusufi}}, \bibinfo {author} {\bibfnamefont {M.~C.}\ \bibnamefont {Werner}}, \bibinfo {author} {\bibfnamefont {A.}~\bibnamefont {Banerjee}}, \ and\ \bibinfo {author} {\bibfnamefont {A.}~\bibnamefont {\"Ovg\"un}},\ }\href {\doibase 10.1103/PhysRevD.95.104012} {\bibfield  {journal} {\bibinfo  {journal} {Phys. Rev. D}\ }\textbf {\bibinfo {volume} {95}},\ \bibinfo {pages} {104012} (\bibinfo {year} {2017})}\BibitemShut {NoStop}%
\bibitem [{\citenamefont {Jusufi}\ and\ \citenamefont {\"Ovg\"un}(2018)}]{PhysRevD.97.024042}%
  \BibitemOpen
  \bibfield  {author} {\bibinfo {author} {\bibfnamefont {K.}~\bibnamefont {Jusufi}}\ and\ \bibinfo {author} {\bibfnamefont {A.}~\bibnamefont {\"Ovg\"un}},\ }\href {\doibase 10.1103/PhysRevD.97.024042} {\bibfield  {journal} {\bibinfo  {journal} {Phys. Rev. D}\ }\textbf {\bibinfo {volume} {97}},\ \bibinfo {pages} {024042} (\bibinfo {year} {2018})}\BibitemShut {NoStop}%
\bibitem [{\citenamefont {\"Ovg\"un}(2018)}]{PhysRevD.98.044033}%
  \BibitemOpen
  \bibfield  {author} {\bibinfo {author} {\bibfnamefont {A.}~\bibnamefont {\"Ovg\"un}},\ }\href {\doibase 10.1103/PhysRevD.98.044033} {\bibfield  {journal} {\bibinfo  {journal} {Phys. Rev. D}\ }\textbf {\bibinfo {volume} {98}},\ \bibinfo {pages} {044033} (\bibinfo {year} {2018})}\BibitemShut {NoStop}%
\bibitem [{\citenamefont {Li}\ and\ \citenamefont {\"Ovg\"un}(2020)}]{PhysRevD.101.024040}%
  \BibitemOpen
  \bibfield  {author} {\bibinfo {author} {\bibfnamefont {Z.}~\bibnamefont {Li}}\ and\ \bibinfo {author} {\bibfnamefont {A.}~\bibnamefont {\"Ovg\"un}},\ }\href {\doibase 10.1103/PhysRevD.101.024040} {\bibfield  {journal} {\bibinfo  {journal} {Phys. Rev. D}\ }\textbf {\bibinfo {volume} {101}},\ \bibinfo {pages} {024040} (\bibinfo {year} {2020})}\BibitemShut {NoStop}%
\bibitem [{\citenamefont {Bozza}(2002)}]{bozza2002gravitational}%
  \BibitemOpen
  \bibfield  {author} {\bibinfo {author} {\bibfnamefont {V.}~\bibnamefont {Bozza}},\ }\href@noop {} {\bibfield  {journal} {\bibinfo  {journal} {Physical Review D}\ }\textbf {\bibinfo {volume} {66}},\ \bibinfo {pages} {103001} (\bibinfo {year} {2002})}\BibitemShut {NoStop}%
\bibitem [{\citenamefont {Tsukamoto}(2017)}]{tsukamoto2017deflection}%
  \BibitemOpen
  \bibfield  {author} {\bibinfo {author} {\bibfnamefont {N.}~\bibnamefont {Tsukamoto}},\ }\href@noop {} {\bibfield  {journal} {\bibinfo  {journal} {Physical Review D}\ }\textbf {\bibinfo {volume} {95}},\ \bibinfo {pages} {064035} (\bibinfo {year} {2017})}\BibitemShut {NoStop}%
\bibitem [{\citenamefont {Mushtaq}\ \emph {et~al.}(2025)\citenamefont {Mushtaq}, \citenamefont {Tiecheng}, \citenamefont {Ditta}, \citenamefont {Mustafa},\ and\ \citenamefont {Maurya}}]{Mushtaq:2024tmp}%
  \BibitemOpen
  \bibfield  {author} {\bibinfo {author} {\bibfnamefont {F.}~\bibnamefont {Mushtaq}}, \bibinfo {author} {\bibfnamefont {X.}~\bibnamefont {Tiecheng}}, \bibinfo {author} {\bibfnamefont {A.}~\bibnamefont {Ditta}}, \bibinfo {author} {\bibfnamefont {G.}~\bibnamefont {Mustafa}}, \ and\ \bibinfo {author} {\bibfnamefont {S.~K.}\ \bibnamefont {Maurya}},\ }\href {\doibase 10.1088/1572-9494/ad7c36} {\bibfield  {journal} {\bibinfo  {journal} {Commun. Theor. Phys.}\ }\textbf {\bibinfo {volume} {77}},\ \bibinfo {pages} {025402} (\bibinfo {year} {2025})}\BibitemShut {NoStop}%
\bibitem [{\citenamefont {Mustafa}\ \emph {et~al.}(2025)\citenamefont {Mustafa}, \citenamefont {Maurya}, \citenamefont {Channuie}, \citenamefont {Bouzenada}, \citenamefont {Abd-Elmonem},\ and\ \citenamefont {Alhubieshi}}]{Mustafa:2024mvx}%
  \BibitemOpen
  \bibfield  {author} {\bibinfo {author} {\bibfnamefont {G.}~\bibnamefont {Mustafa}}, \bibinfo {author} {\bibfnamefont {S.~K.}\ \bibnamefont {Maurya}}, \bibinfo {author} {\bibfnamefont {P.}~\bibnamefont {Channuie}}, \bibinfo {author} {\bibfnamefont {A.}~\bibnamefont {Bouzenada}}, \bibinfo {author} {\bibfnamefont {A.}~\bibnamefont {Abd-Elmonem}}, \ and\ \bibinfo {author} {\bibfnamefont {N.}~\bibnamefont {Alhubieshi}},\ }\href {\doibase 10.1016/j.dark.2024.101753} {\bibfield  {journal} {\bibinfo  {journal} {Phys. Dark Univ.}\ }\textbf {\bibinfo {volume} {47}},\ \bibinfo {pages} {101753} (\bibinfo {year} {2025})}\BibitemShut {NoStop}%
\bibitem [{\citenamefont {Wang}\ and\ \citenamefont {Battista}(2025)}]{Wang:2025fmz}%
  \BibitemOpen
  \bibfield  {author} {\bibinfo {author} {\bibfnamefont {Z.-L.}\ \bibnamefont {Wang}}\ and\ \bibinfo {author} {\bibfnamefont {E.}~\bibnamefont {Battista}},\ }\href {\doibase 10.1140/epjc/s10052-025-13833-7} {\bibfield  {journal} {\bibinfo  {journal} {Eur. Phys. J. C}\ }\textbf {\bibinfo {volume} {85}},\ \bibinfo {pages} {304} (\bibinfo {year} {2025})},\ \Eprint {http://arxiv.org/abs/2501.14516} {arXiv:2501.14516 [gr-qc]} \BibitemShut {NoStop}%
\bibitem [{\citenamefont {Capozziello}\ \emph {et~al.}(2024)\citenamefont {Capozziello}, \citenamefont {De~Bianchi},\ and\ \citenamefont {Battista}}]{Capozziello:2024ucm}%
  \BibitemOpen
  \bibfield  {author} {\bibinfo {author} {\bibfnamefont {S.}~\bibnamefont {Capozziello}}, \bibinfo {author} {\bibfnamefont {S.}~\bibnamefont {De~Bianchi}}, \ and\ \bibinfo {author} {\bibfnamefont {E.}~\bibnamefont {Battista}},\ }\href {\doibase 10.1103/PhysRevD.109.104060} {\bibfield  {journal} {\bibinfo  {journal} {Phys. Rev. D}\ }\textbf {\bibinfo {volume} {109}},\ \bibinfo {pages} {104060} (\bibinfo {year} {2024})},\ \Eprint {http://arxiv.org/abs/2404.17267} {arXiv:2404.17267 [gr-qc]} \BibitemShut {NoStop}%
\bibitem [{\citenamefont {Sekhmani}\ \emph {et~al.}(2025)\citenamefont {Sekhmani}, \citenamefont {Maurya}, \citenamefont {Jasim}, \citenamefont {Sakalli}, \citenamefont {Rayimbaev},\ and\ \citenamefont {Ibragimov}}]{Sekhmani:2025kav}%
  \BibitemOpen
  \bibfield  {author} {\bibinfo {author} {\bibfnamefont {Y.}~\bibnamefont {Sekhmani}}, \bibinfo {author} {\bibfnamefont {S.~K.}\ \bibnamefont {Maurya}}, \bibinfo {author} {\bibfnamefont {M.~K.}\ \bibnamefont {Jasim}}, \bibinfo {author} {\bibfnamefont {I.}~\bibnamefont {Sakalli}}, \bibinfo {author} {\bibfnamefont {J.}~\bibnamefont {Rayimbaev}}, \ and\ \bibinfo {author} {\bibfnamefont {I.}~\bibnamefont {Ibragimov}},\ }\href {\doibase 10.1140/epjc/s10052-025-13932-5} {\bibfield  {journal} {\bibinfo  {journal} {Eur. Phys. J. C}\ }\textbf {\bibinfo {volume} {85}},\ \bibinfo {pages} {229} (\bibinfo {year} {2025})}\BibitemShut {NoStop}%
\bibitem [{\citenamefont {Javed}\ \emph {et~al.}(2024)\citenamefont {Javed}, \citenamefont {Mustafa}, \citenamefont {Fatima}, \citenamefont {Maurya}, \citenamefont {Alshehri},\ and\ \citenamefont {Mubeen}}]{Javed:2024nnt}%
  \BibitemOpen
  \bibfield  {author} {\bibinfo {author} {\bibfnamefont {F.}~\bibnamefont {Javed}}, \bibinfo {author} {\bibfnamefont {G.}~\bibnamefont {Mustafa}}, \bibinfo {author} {\bibfnamefont {G.}~\bibnamefont {Fatima}}, \bibinfo {author} {\bibfnamefont {S.~K.}\ \bibnamefont {Maurya}}, \bibinfo {author} {\bibfnamefont {M.~H.}\ \bibnamefont {Alshehri}}, \ and\ \bibinfo {author} {\bibfnamefont {I.}~\bibnamefont {Mubeen}},\ }\href {\doibase 10.1016/j.jheap.2024.09.003} {\bibfield  {journal} {\bibinfo  {journal} {JHEAp}\ }\textbf {\bibinfo {volume} {44}},\ \bibinfo {pages} {60} (\bibinfo {year} {2024})}\BibitemShut {NoStop}%
\bibitem [{\citenamefont {Ashraf}\ \emph {et~al.}(2025)\citenamefont {Ashraf}, \citenamefont {Javed}, \citenamefont {Maurya}, \citenamefont {Channuie}, \citenamefont {Cilli},\ and\ \citenamefont {G\"udekli}}]{Ashraf:2025edl}%
  \BibitemOpen
  \bibfield  {author} {\bibinfo {author} {\bibfnamefont {A.}~\bibnamefont {Ashraf}}, \bibinfo {author} {\bibfnamefont {F.}~\bibnamefont {Javed}}, \bibinfo {author} {\bibfnamefont {S.~K.}\ \bibnamefont {Maurya}}, \bibinfo {author} {\bibfnamefont {P.}~\bibnamefont {Channuie}}, \bibinfo {author} {\bibfnamefont {A.}~\bibnamefont {Cilli}}, \ and\ \bibinfo {author} {\bibfnamefont {E.}~\bibnamefont {G\"udekli}},\ }\href {\doibase 10.1016/j.dark.2025.101853} {\bibfield  {journal} {\bibinfo  {journal} {Phys. Dark Univ.}\ }\textbf {\bibinfo {volume} {48}},\ \bibinfo {pages} {101853} (\bibinfo {year} {2025})}\BibitemShut {NoStop}%
\bibitem [{\citenamefont {Poulis}\ and\ \citenamefont {Soares}(2022)}]{poulis2022exact}%
  \BibitemOpen
  \bibfield  {author} {\bibinfo {author} {\bibfnamefont {F.}~\bibnamefont {Poulis}}\ and\ \bibinfo {author} {\bibfnamefont {M.}~\bibnamefont {Soares}},\ }\href@noop {} {\bibfield  {journal} {\bibinfo  {journal} {The European Physical Journal C}\ }\textbf {\bibinfo {volume} {82}},\ \bibinfo {pages} {1} (\bibinfo {year} {2022})}\BibitemShut {NoStop}%
\bibitem [{\citenamefont {Kosteleck\'y}\ and\ \citenamefont {Li}(2021)}]{Kostelecky:2020hbb}%
  \BibitemOpen
  \bibfield  {author} {\bibinfo {author} {\bibfnamefont {V.~A.}\ \bibnamefont {Kosteleck\'y}}\ and\ \bibinfo {author} {\bibfnamefont {Z.}~\bibnamefont {Li}},\ }\href {\doibase 10.1103/PhysRevD.103.024059} {\bibfield  {journal} {\bibinfo  {journal} {Phys. Rev. D}\ }\textbf {\bibinfo {volume} {103}},\ \bibinfo {pages} {024059} (\bibinfo {year} {2021})},\ \Eprint {http://arxiv.org/abs/2008.12206} {arXiv:2008.12206 [gr-qc]} \BibitemShut {NoStop}%
\bibitem [{\citenamefont {Bertolami}\ and\ \citenamefont {Paramos}(2005)}]{Bertolami:2005bh}%
  \BibitemOpen
  \bibfield  {author} {\bibinfo {author} {\bibfnamefont {O.}~\bibnamefont {Bertolami}}\ and\ \bibinfo {author} {\bibfnamefont {J.}~\bibnamefont {Paramos}},\ }\href {\doibase 10.1103/PhysRevD.72.044001} {\bibfield  {journal} {\bibinfo  {journal} {Phys. Rev. D}\ }\textbf {\bibinfo {volume} {72}},\ \bibinfo {pages} {044001} (\bibinfo {year} {2005})},\ \Eprint {http://arxiv.org/abs/hep-th/0504215} {arXiv:hep-th/0504215} \BibitemShut {NoStop}%
\bibitem [{\citenamefont {Casana}\ \emph {et~al.}(2018)\citenamefont {Casana}, \citenamefont {Cavalcante}, \citenamefont {Poulis},\ and\ \citenamefont {Santos}}]{Casana:2017jkc}%
  \BibitemOpen
  \bibfield  {author} {\bibinfo {author} {\bibfnamefont {R.}~\bibnamefont {Casana}}, \bibinfo {author} {\bibfnamefont {A.}~\bibnamefont {Cavalcante}}, \bibinfo {author} {\bibfnamefont {F.~P.}\ \bibnamefont {Poulis}}, \ and\ \bibinfo {author} {\bibfnamefont {E.~B.}\ \bibnamefont {Santos}},\ }\href {\doibase 10.1103/PhysRevD.97.104001} {\bibfield  {journal} {\bibinfo  {journal} {Phys. Rev. D}\ }\textbf {\bibinfo {volume} {97}},\ \bibinfo {pages} {104001} (\bibinfo {year} {2018})},\ \Eprint {http://arxiv.org/abs/1711.02273} {arXiv:1711.02273 [gr-qc]} \BibitemShut {NoStop}%
\bibitem [{\citenamefont {Perlick}\ and\ \citenamefont {Tsupko}(2022)}]{Perlick:2021aok}%
  \BibitemOpen
  \bibfield  {author} {\bibinfo {author} {\bibfnamefont {V.}~\bibnamefont {Perlick}}\ and\ \bibinfo {author} {\bibfnamefont {O.~Y.}\ \bibnamefont {Tsupko}},\ }\href {\doibase 10.1016/j.physrep.2021.10.004} {\bibfield  {journal} {\bibinfo  {journal} {Phys. Rept.}\ }\textbf {\bibinfo {volume} {947}},\ \bibinfo {pages} {1} (\bibinfo {year} {2022})},\ \Eprint {http://arxiv.org/abs/2105.07101} {arXiv:2105.07101 [gr-qc]} \BibitemShut {NoStop}%
\bibitem [{\citenamefont {Vagnozzi}\ \emph {et~al.}(2023)\citenamefont {Vagnozzi} \emph {et~al.}}]{Vagnozzi:2022moj}%
  \BibitemOpen
  \bibfield  {author} {\bibinfo {author} {\bibfnamefont {S.}~\bibnamefont {Vagnozzi}} \emph {et~al.},\ }\href {\doibase 10.1088/1361-6382/acd97b} {\bibfield  {journal} {\bibinfo  {journal} {Class. Quant. Grav.}\ }\textbf {\bibinfo {volume} {40}},\ \bibinfo {pages} {165007} (\bibinfo {year} {2023})},\ \Eprint {http://arxiv.org/abs/2205.07787} {arXiv:2205.07787 [gr-qc]} \BibitemShut {NoStop}%
\bibitem [{\citenamefont {Joshi}\ \emph {et~al.}(2020)\citenamefont {Joshi}, \citenamefont {Dey}, \citenamefont {Joshi},\ and\ \citenamefont {Bambhaniya}}]{Joshi:2020tlq}%
  \BibitemOpen
  \bibfield  {author} {\bibinfo {author} {\bibfnamefont {A.~B.}\ \bibnamefont {Joshi}}, \bibinfo {author} {\bibfnamefont {D.}~\bibnamefont {Dey}}, \bibinfo {author} {\bibfnamefont {P.~S.}\ \bibnamefont {Joshi}}, \ and\ \bibinfo {author} {\bibfnamefont {P.}~\bibnamefont {Bambhaniya}},\ }\href {\doibase 10.1103/PhysRevD.102.024022} {\bibfield  {journal} {\bibinfo  {journal} {Phys. Rev. D}\ }\textbf {\bibinfo {volume} {102}},\ \bibinfo {pages} {024022} (\bibinfo {year} {2020})},\ \Eprint {http://arxiv.org/abs/2004.06525} {arXiv:2004.06525 [gr-qc]} \BibitemShut {NoStop}%
\bibitem [{\citenamefont {Psaltis}(2008)}]{Psaltis:2007rv}%
  \BibitemOpen
  \bibfield  {author} {\bibinfo {author} {\bibfnamefont {D.}~\bibnamefont {Psaltis}},\ }\href {\doibase 10.1103/PhysRevD.77.064006} {\bibfield  {journal} {\bibinfo  {journal} {Phys. Rev. D}\ }\textbf {\bibinfo {volume} {77}},\ \bibinfo {pages} {064006} (\bibinfo {year} {2008})},\ \Eprint {http://arxiv.org/abs/0704.2426} {arXiv:0704.2426 [astro-ph]} \BibitemShut {NoStop}%
\bibitem [{\citenamefont {Cunha}\ and\ \citenamefont {Herdeiro}(2018)}]{Cunha:2018acu}%
  \BibitemOpen
  \bibfield  {author} {\bibinfo {author} {\bibfnamefont {P.~V.~P.}\ \bibnamefont {Cunha}}\ and\ \bibinfo {author} {\bibfnamefont {C.~A.~R.}\ \bibnamefont {Herdeiro}},\ }\href {\doibase 10.1007/s10714-018-2361-9} {\bibfield  {journal} {\bibinfo  {journal} {Gen. Rel. Grav.}\ }\textbf {\bibinfo {volume} {50}},\ \bibinfo {pages} {42} (\bibinfo {year} {2018})},\ \Eprint {http://arxiv.org/abs/1801.00860} {arXiv:1801.00860 [gr-qc]} \BibitemShut {NoStop}%
\bibitem [{\citenamefont {Dokuchaev}\ and\ \citenamefont {Nazarova}(2020)}]{Dokuchaev:2019jqq}%
  \BibitemOpen
  \bibfield  {author} {\bibinfo {author} {\bibfnamefont {V.~I.}\ \bibnamefont {Dokuchaev}}\ and\ \bibinfo {author} {\bibfnamefont {N.~O.}\ \bibnamefont {Nazarova}},\ }\href {\doibase 10.3367/UFNe.2020.01.038717} {\bibfield  {journal} {\bibinfo  {journal} {Usp. Fiz. Nauk}\ }\textbf {\bibinfo {volume} {190}},\ \bibinfo {pages} {627} (\bibinfo {year} {2020})},\ \Eprint {http://arxiv.org/abs/1911.07695} {arXiv:1911.07695 [gr-qc]} \BibitemShut {NoStop}%
\bibitem [{\citenamefont {Psaltis}\ \emph {et~al.}(2020)\citenamefont {Psaltis} \emph {et~al.}}]{EventHorizonTelescope:2020qrl}%
  \BibitemOpen
  \bibfield  {author} {\bibinfo {author} {\bibfnamefont {D.}~\bibnamefont {Psaltis}} \emph {et~al.} (\bibinfo {collaboration} {Event Horizon Telescope}),\ }\href {\doibase 10.1103/PhysRevLett.125.141104} {\bibfield  {journal} {\bibinfo  {journal} {Phys. Rev. Lett.}\ }\textbf {\bibinfo {volume} {125}},\ \bibinfo {pages} {141104} (\bibinfo {year} {2020})},\ \Eprint {http://arxiv.org/abs/2010.01055} {arXiv:2010.01055 [gr-qc]} \BibitemShut {NoStop}%
\bibitem [{\citenamefont {Akiyama}\ \emph {et~al.}(2022{\natexlab{b}})\citenamefont {Akiyama} \emph {et~al.}}]{EventHorizonTelescope:2022wkp}%
  \BibitemOpen
  \bibfield  {author} {\bibinfo {author} {\bibfnamefont {K.}~\bibnamefont {Akiyama}} \emph {et~al.} (\bibinfo {collaboration} {Event Horizon Telescope}),\ }\href {\doibase 10.3847/2041-8213/ac6674} {\bibfield  {journal} {\bibinfo  {journal} {Astrophys. J. Lett.}\ }\textbf {\bibinfo {volume} {930}},\ \bibinfo {pages} {L12} (\bibinfo {year} {2022}{\natexlab{b}})},\ \Eprint {http://arxiv.org/abs/2311.08680} {arXiv:2311.08680 [astro-ph.HE]} \BibitemShut {NoStop}%
\bibitem [{\citenamefont {Huang}\ \emph {et~al.}(2024)\citenamefont {Huang}, \citenamefont {Cao},\ and\ \citenamefont {Lu}}]{Huang:2023bto}%
  \BibitemOpen
  \bibfield  {author} {\bibinfo {author} {\bibfnamefont {Y.}~\bibnamefont {Huang}}, \bibinfo {author} {\bibfnamefont {Z.}~\bibnamefont {Cao}}, \ and\ \bibinfo {author} {\bibfnamefont {Z.}~\bibnamefont {Lu}},\ }\href {\doibase 10.1088/1475-7516/2024/01/013} {\bibfield  {journal} {\bibinfo  {journal} {JCAP}\ }\textbf {\bibinfo {volume} {01}},\ \bibinfo {pages} {013} (\bibinfo {year} {2024})},\ \Eprint {http://arxiv.org/abs/2306.04145} {arXiv:2306.04145 [gr-qc]} \BibitemShut {NoStop}%
\bibitem [{\citenamefont {Huang}\ and\ \citenamefont {Cao}(2023)}]{Huang:2022iwl}%
  \BibitemOpen
  \bibfield  {author} {\bibinfo {author} {\bibfnamefont {Y.}~\bibnamefont {Huang}}\ and\ \bibinfo {author} {\bibfnamefont {Z.}~\bibnamefont {Cao}},\ }\href {\doibase 10.1140/epjc/s10052-023-11180-z} {\bibfield  {journal} {\bibinfo  {journal} {Eur. Phys. J. C}\ }\textbf {\bibinfo {volume} {83}},\ \bibinfo {pages} {80} (\bibinfo {year} {2023})},\ \Eprint {http://arxiv.org/abs/2212.04254} {arXiv:2212.04254 [gr-qc]} \BibitemShut {NoStop}%
\bibitem [{\citenamefont {Ono}\ \emph {et~al.}(2017)\citenamefont {Ono}, \citenamefont {Ishihara},\ and\ \citenamefont {Asada}}]{PhysRevD.96.104037}%
  \BibitemOpen
  \bibfield  {author} {\bibinfo {author} {\bibfnamefont {T.}~\bibnamefont {Ono}}, \bibinfo {author} {\bibfnamefont {A.}~\bibnamefont {Ishihara}}, \ and\ \bibinfo {author} {\bibfnamefont {H.}~\bibnamefont {Asada}},\ }\href {\doibase 10.1103/PhysRevD.96.104037} {\bibfield  {journal} {\bibinfo  {journal} {Phys. Rev. D}\ }\textbf {\bibinfo {volume} {96}},\ \bibinfo {pages} {104037} (\bibinfo {year} {2017})}\BibitemShut {NoStop}%
\bibitem [{\citenamefont {Filho}\ \emph {et~al.}(2023)\citenamefont {Filho}, \citenamefont {Nascimento}, \citenamefont {Petrov},\ and\ \citenamefont {Porf\'\i{}rio}}]{Filho:2022yrk}%
  \BibitemOpen
  \bibfield  {author} {\bibinfo {author} {\bibfnamefont {A.~A.~A.}\ \bibnamefont {Filho}}, \bibinfo {author} {\bibfnamefont {J.~R.}\ \bibnamefont {Nascimento}}, \bibinfo {author} {\bibfnamefont {A.~Y.}\ \bibnamefont {Petrov}}, \ and\ \bibinfo {author} {\bibfnamefont {P.~J.}\ \bibnamefont {Porf\'\i{}rio}},\ }\href {\doibase 10.1103/PhysRevD.108.085010} {\bibfield  {journal} {\bibinfo  {journal} {Phys. Rev. D}\ }\textbf {\bibinfo {volume} {108}},\ \bibinfo {pages} {085010} (\bibinfo {year} {2023})},\ \Eprint {http://arxiv.org/abs/2211.11821} {arXiv:2211.11821 [gr-qc]} \BibitemShut {NoStop}%
\bibitem [{\citenamefont {Maluf}\ and\ \citenamefont {Neves}(2021{\natexlab{b}})}]{Maluf:2021lwh}%
  \BibitemOpen
  \bibfield  {author} {\bibinfo {author} {\bibfnamefont {R.~V.}\ \bibnamefont {Maluf}}\ and\ \bibinfo {author} {\bibfnamefont {J.~C.~S.}\ \bibnamefont {Neves}},\ }\href {\doibase 10.1088/1475-7516/2021/10/038} {\bibfield  {journal} {\bibinfo  {journal} {JCAP}\ }\textbf {\bibinfo {volume} {10}},\ \bibinfo {pages} {038} (\bibinfo {year} {2021}{\natexlab{b}})},\ \Eprint {http://arxiv.org/abs/2105.08659} {arXiv:2105.08659 [gr-qc]} \BibitemShut {NoStop}%
\end{thebibliography}%

\end{document}